\documentclass[10pt,final]{iopart}

\expandafter\let\csname equation*\endcsname\relax
\expandafter\let\csname endequation*\endcsname\relax
\usepackage{amsmath}

\usepackage{amssymb}

\usepackage[makeroom]{cancel}

\usepackage{mhchem}

\usepackage{graphicx}

\usepackage{dcolumn}

\usepackage{bm}
\usepackage[normalem]{ulem}
\usepackage{color}

\usepackage{verbatim}

\usepackage{epstopdf}

\newcommand{\ii}{{\rm i}}

\newcommand{\bq}{\mathbf{q}}
\newcommand{\bv}{\mathbf{v}}

\newcommand{\bff}{\mathbf{f}}
\newcommand{\bu}{\mathbf{u}}

\newcommand{\cH}{{\cal H}}
\newcommand{\bw}{\mathbf{w}}
\newcommand{\bbr}{\mathbf{r}}

\newcommand{\tri}{\triangle}

\newcommand{\sep}{ \ \ \ , \ \ \ }

\newcommand{\beq}{\begin{equation}}
\newcommand{\eeq}{\end{equation}}
\newcommand{\beqn}{\begin{eqnarray}}
\newcommand{\eeqn}{\end{eqnarray}}
\newcommand{\pp}{\partial}
\newcommand{\dd}{{\rm d}}
\newcommand{\ee}{{\rm e}}
\newcommand{\eq}{Eq.\ }

\newcommand{\eqs}{Eqs }
\newcommand{\fig}{Fig.\ }

\newcommand{\cO}{{\cal O}}

\newcommand{\Sec}{Sec.\ }

\newcommand{\la}{\langle}

\newcommand{\ra}{\rangle}

\newcommand{\hatPin}{\hat{P}_{\rm in}}
\newcommand{\hatPo}{\hat{P}_{\rm out}}

\newcommand{\Pin}{P_{\rm in}}
\newcommand{\Po}{P_{\rm out}}

\newcommand{\Pinf}{P_{\rm out}(\infty)}
\newcommand{\Pio}{P_{\rm in,out}}
\newcommand{\Sio}{S_{\rm in,out}}

\newcommand{\Sin}{S_{\rm in}}
\newcommand{\So}{S_{\rm out}}

\newcommand{\Jio}{J_{\rm in \rightarrow out}}
\newcommand{\Joi}{J_{\rm out \rightarrow in}}

\newcommand{\Ptot}{P_{\rm tot}}

\newcommand{\cblack}{\color{black}}
\newcommand{\cblue}{\color{blue}}

\newcommand{\Eqref}[1]{Eq.\ \eqref{#1}}

\newcommand{\GT}{\eqs \eqref{gtIn} and \eqref{gtP}}

\begin{document}

\title{Novel physics arising from phase transitions in biology}

\author{Chiu Fan Lee}
\address{Department of Bioengineering, Imperial College London, South Kensington Campus, London SW7 2AZ, U.K.}
\ead{c.lee@imperial.ac.uk}
\author{Jean David Wurtz}
\address{Department of Bioengineering, Imperial College London, South Kensington Campus, London SW7 2AZ, U.K.}
\begin{abstract}
	Phase transitions, such as the freezing of water and the magnetisation of a ferromagnet upon lowering the ambient temperature, are familiar physical phenomena. Interestingly, such a collective change of behaviour at a phase transition is also of importance to living systems. From cytoplasmic organisation inside a cell to the collective migration of cell tissue during organismal development and wound healing, phase transitions have emerged as  key mechanisms underlying many crucial biological processes. However, a living system is fundamentally different from a thermal system, with driven chemical reactions (e.g., metabolism)  and motility being two hallmarks of its non-equilibrium nature. In this review, we 
	will discuss how driven chemical reactions can arrest universal coarsening kinetics expected from thermal phase separation, and how motility leads to the emergence of a novel universality class when the rotational symmetry is spontaneously broken in  an incompressible fluid.
\end{abstract}
\submitto{\JPD}
\maketitle

%

\section{Introduction}
\begin{figure}[]
	\centering
	\includegraphics[scale=.6]{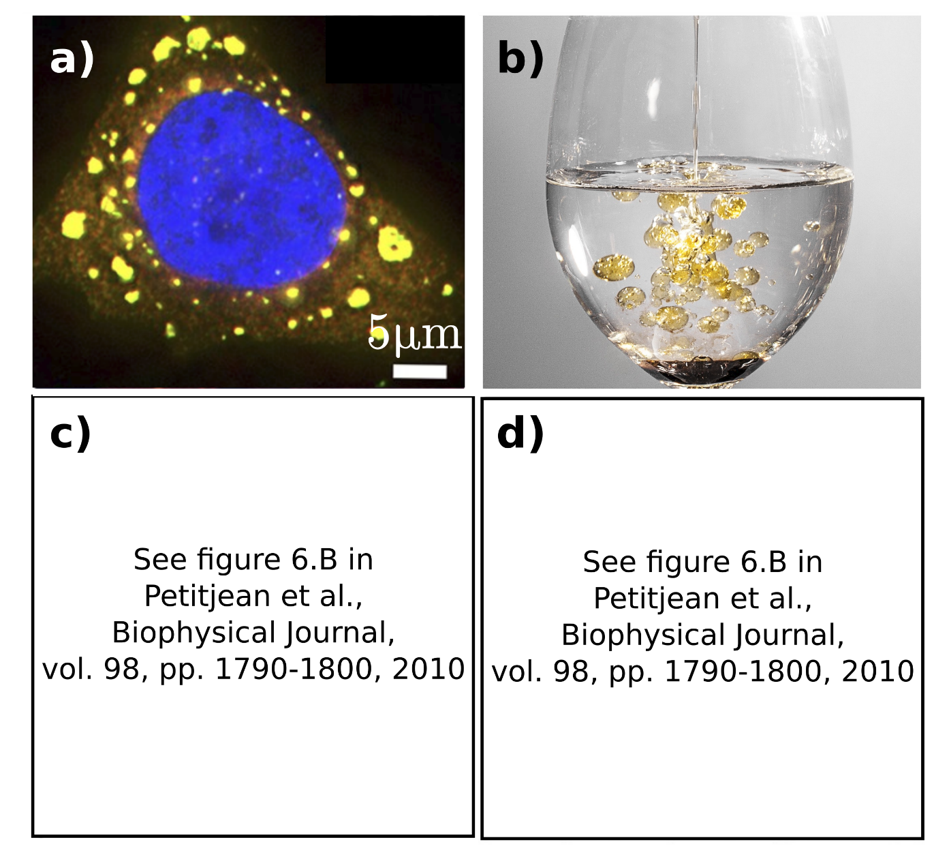}
	\caption{
		{\it Cytoplasmic phase separation and tissue dynamics as active matter.}
		a) In many distinct types of cells, certain proteins can phase separate from the cytosol to assemble membrane-less organelles, such as the stress granules (yellow drops) shown here in human epithelial cells (\textbf{HeLa}) \cite{wheeler_elife16}, akin to oil drop formation in an oil-water mixture (b). c) In a monolayer of Madin-Darby Canine Kidney ({\bf MDCK}) cells, the cells in the tissue can undergo dynamical rearrangement as shown by the snapshot of the velocity field shown in (d) \cite{petitjean_biophysj10}. Figure a) is adapted from [J.R. Wheeler et al., eLife vol. 5, pp. e18413, 2018], licensed under CC BY 4.0. Figure b) is licensed under CC0 Creative Commons, Pixabay.com. 
		\label{fig:mainfig}
	}
\end{figure}

Collective phenomena are intimately linked to the phenomenon of phase transitions in physics. At a typical phase transition, a many-body system with constituents that interact only locally with their neighbours, be they molecules or living  organisms, can collectively change their behaviour upon a subtle change of a single parameter, to the extent that the qualitative behaviour of the whole system is modified. Phase transitions encompass many everyday phenomena such as oil drop formation in a salad dressing and  magnetisation in some metals. 
The study of phase transitions is of fundamental interest to physicists because of the emergence of  {\it universal}  behaviours at a phase transition. By universal behaviour, we mean that certain properties of the system are highly independent of the system's microscopic details. In the salad dressing example,  such property can be the power law exponent that governs how the average size of oil drops changes with time;  in the example of magnetisation, it can be the power law exponent that governs how the  correlation function of two atomic spins  decays with respect to 
their distance. Recently, phase transitions in living systems have also been under intense attention. Indeed, the generic non-equilibrium nature of biological systems have given rise to novel universal behaviours not seen before.   In this review, we will focus on two such examples: phase separation with driven chemical reactions, motivated by the mechanism underlying the formation of 
some membrane-less organelles in cells \cite{Berry2018,hyman_annrev14}, and spontaneous symmetry breaking in incompressible active matter, motivated by its relevance to biological tissues \cite{haga_biophysj05,szabo_pre06,Giavazzi2018}  (\fig \ref{fig:mainfig}).

In Sect.~2, we will first describe the relevance of phase separation in cytoplasmic organisation and then review the latest findings on how driven chemical reactions (e.g., adenosine triphosphate ({\bf ATP})-driven phosphorylation)   can lead to co-existing phase-separated protein drops in the cytoplasm, contrary to the universal coarsening behaviour expected in its equilibrium counter part. In Sect.~3, motivated by the collective behaviour  found in motile organisms, we will introduce a generic model of incompressible active fluids from a symmetry consideration. We will then elucidate how a novel critical behaviour emerges at the onset of collective motion, and discuss the universal behaviour of a two dimensional incompressible active fluid in the ordered phase. Finally, we will end with Conclusion \& Outlook.

\section{Non-equilibrium phase separation: a mechanism for cytoplasmic organisation}

\subsection{Membrane-less organelles}

Biological cells organise their contents in distinct compartments called organelles, typically enclosed by a lipid membrane that forms a physical barrier and controls molecular exchanges with the surrounding cytosol. Recently an intriguing class of organelles lacking a membrane is being studied intensely \cite{brangwynne_softmatt11}. Membrane-less organelles have attracted an intense interest from the biology community as they are present in many organisms from yeast to mammal cells, and are critical for multiple biological functions. For example P granules are involved in the asymmetric division of the \textit{Caenorhabditis elegans} embryo \cite{updike_andrology10}, and stress granules assemble during environmental stress and protect cytoplasmic RNA from degradation \cite{protter_trends16} (\fig \ref{fig:mainfig} a)). Membrane-less organelles are generally spherical,  fuse together upon contact \cite{brangwynne_jcb13, brangwynne_science09}, and their components quickly shuttle in and out \cite{kedersha_jcb00, bley_17}, thus resembling liquid drops.
Indeed,  strong experimental evidence indicates that membrane-less organelles are assembled via liquid-liquid phase 
separation \cite{hyman_annrev14, saha_cell16, molliex_cell15}, a common phenomenon in every day life responsible for example for oil drop formation in water (\fig \ref{fig:mainfig} b).  Under the equilibrium condition phase separation is well understood \cite{hill_book86}. However cells are driven away from equilibrium by multiple energy-consuming processes such as ATP-driven protein phosphorylation \cite{alberts_2008}, which can potentially affect the phase-separating behavior of membrane-less constituents. For example P granules do not distribute homogeneously in the cytoplasm but preferentially to the posterior side of the cell \cite{lee_prl13}, and stress granules form and dissolve according to environmental cues \cite{gilks_mboc04}. The fascinating physics associated to membrane-less organelles are only beginning to be investigated \cite{brangwynne_science09,brangwynne_pnas11,lee_prl13,brangwynne_jcb13,zwicker_pnas14,zwicker_pre15,weber_njp17,weber_r18}.

In this section, we will start with a brief summary of relevant principles of equilibrium phase separation in \Sec \ref{sec:equi}.  We will then review the latest progress on  phase separation driven out of equilibrium by energy-driven chemical reactions in \Sec \ref{sec:noneq}. Specifically we will  focus on  a ternary fluid model of the cell cytoplasm where chemical reactions can convert phase-separating molecules into soluble molecules and \textit{vice versa}. We will show how such reactions can control drops assembly and size, and suppress Ostwald ripening, allowing a collection of organelles to coexist in the cytoplasm.

\subsection{ Equilibrium phase separation}
\label{sec:equi}

\begin{figure}[]
	\centering
	\includegraphics[scale=1]{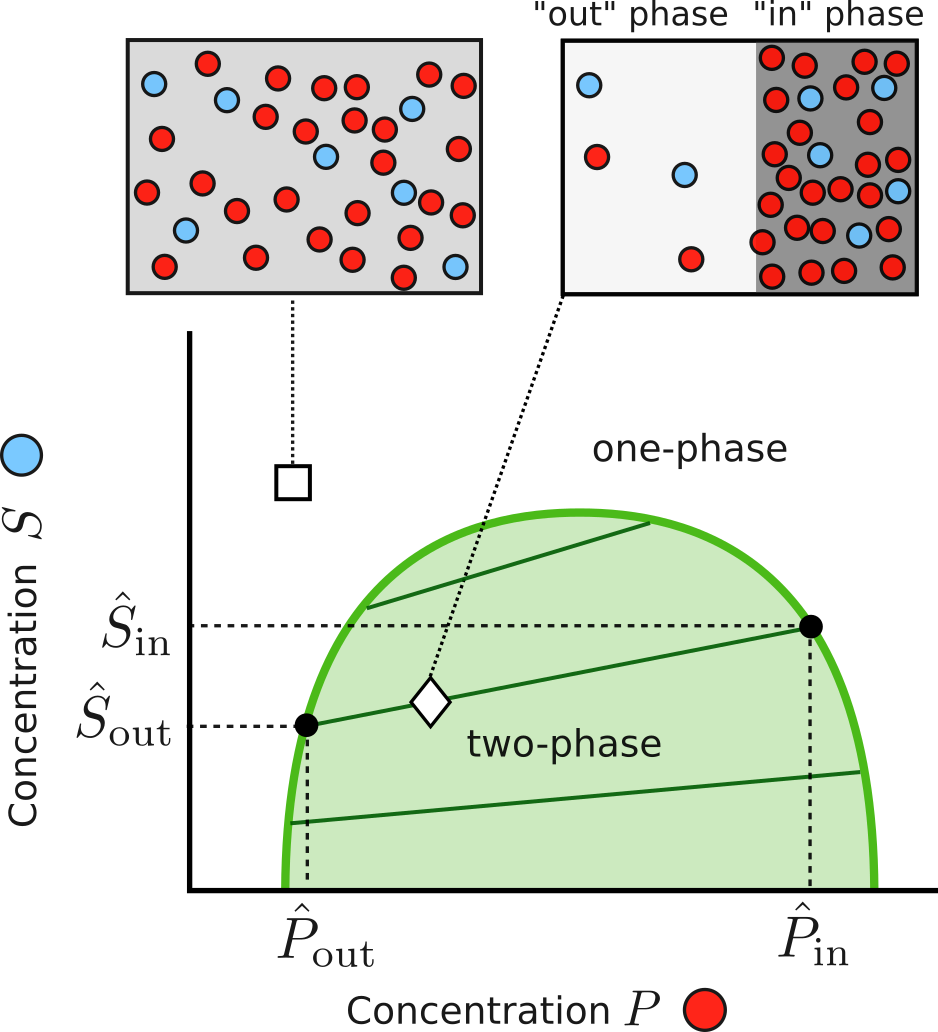}
	\caption{
		\textit{Equilibrium phase diagram of a ternary mixture} composed of molecules $P$ (red disks), $S$ (blue disks) and $C$ (not shown). Outside the phase boundary (green curve) the system is homogeneous (``$\square$" symbol). Inside the phase boundary (``$\diamondsuit$" symbol) the system phase separates into two phases ``in" and ``out" of distinct concentrations. The coexistence concentrations $\hat P_{\rm in,out},\hat S_{\rm in,out}$ are given by the intersections between the tie-lines (straight lines) and the phase boundary.
		\label{fig:equi1}
	}
	
\end{figure}

Interactions between molecules can cause a homogeneous system to undergo a phase separation, i.e. the spontaneous partitioning of a system into multiple phases of distinct properties such as concentration \cite{hill_book86}. The transition from the homogeneous state to the phase-separated state is controlled by parameters such as temperature, pressure or concentrations. The set of parameters leading to phase separation are represented in a phase diagram, as shown in \fig \ref{fig:equi1} for a ternary mixture composed of molecules $P$ (red disks), $S$ (blue disks) and $C$ (not shown). The molecular concentrations are labelled by the same symbols $P,S,C$. We assume incompressibility and that all three types of molecules occupy the same volume, so the combined concentration $\psi
\equiv P+S+C$ is homogeneous. The concentration $C$ at any point in the phase diagram is therefore given by $ \psi-P-S$. Outside the phase boundary (green curve) the system is homogeneous (``$\square$" symbol). Inside the phase boundary (``$\diamondsuit$" symbol) the system phase separates into two phases (``in" and ``out") of distinct concentrations ($\hat P_{\rm in,out},\hat S_{\rm in,out}$), given by the intersections between the tie-lines (straight lines) and the phase boundary.

At the equilibrium condition a multi-drop system is unstable due to Ostwald ripening that causes large drops to grow and small drops to evaporate \cite{ostwald1897studien, lifshitz_jpcs61}, and coalescence caused by the fusion of drops upon contact \cite{siggia_pra79}. Eventually a unique drop remains in a finite system. Since the crowded environment of the cytoplasm inhibits the diffusion of macromolecular aggregates \cite{weiss_biophysj04} we will ignore drop coalescence in this review and focus on Ostwald ripening.

Ostwald ripening is caused by two ingredients. One is the Gibbs-Thomson relation that relates the coexistence concentration to the drop radius. For example for the $P$ concentration we have:
\beqn
\label{gtIn}
\Pin(R) &=& \hatPin \\
\label{gtP}
\Po(R) &=& \cblack{\hatPo \left( 1 + \frac{l_c}{R} \right) }\ ,
\eeqn
were $l_c$ is a capillary length \cite{weber_r18} and $\hat P_{\rm in,out}$ are the coexistence concentrations for a flat interface ($R\rightarrow \infty$, \fig \ref{fig:equi1}). The smaller the drop, the larger the concentration outside which is a consequence of the Laplace pressure \cite{hill_book86}.

The second ingredient driving Ostwald ripening is the existence of a diffusive concentration profile between drops, which can be approximated by an ideal gas diffusion profile in the case of small concentration outside drops \cite{lifshitz_jpcs61}:
\beqn
\label{diffusionEq}
\frac{\partial \Po(r,t)}{\partial t} = D \nabla^2 \Po(r,t) \ ,
\eeqn
where $\Po(r,t)$ is the $P$ profile outside a drop at time $t$ and distance $r$ from the drop centre, and we have assumed spherical symmetry centred on the drop. $D$ is the molecular diffusion coefficient. We use a crucial assumption known as the \textit{quasi-static approximation}: the dynamics of drop radii is much slower than the equilibration of the concentration profiles. Therefore the profiles can be assumed to be always at steady state, and imposing the Gibbs-Thomson relation at the interface (\Eqref{gtP}) we find:
\beqn
\label{profileEq}
\Po(r)=\Po(\infty) - \frac{R}{r}\left(\Delta-\frac{\hatPo l_c}{R}\right) \ ,
\eeqn
where $\Pinf$ is the concentration far from drops and $\Delta \equiv \Pinf - \hatPo$ is commonly referred to as the supersaturation. We have also assumed small drop density so that drops are on average far from each other and the concentration far from drops $\Po(\infty)$ is homogeneous. In other words, drops interact with each other only via this common far-field.

\begin{figure}[h!]
	\centering
	\includegraphics[scale=1]{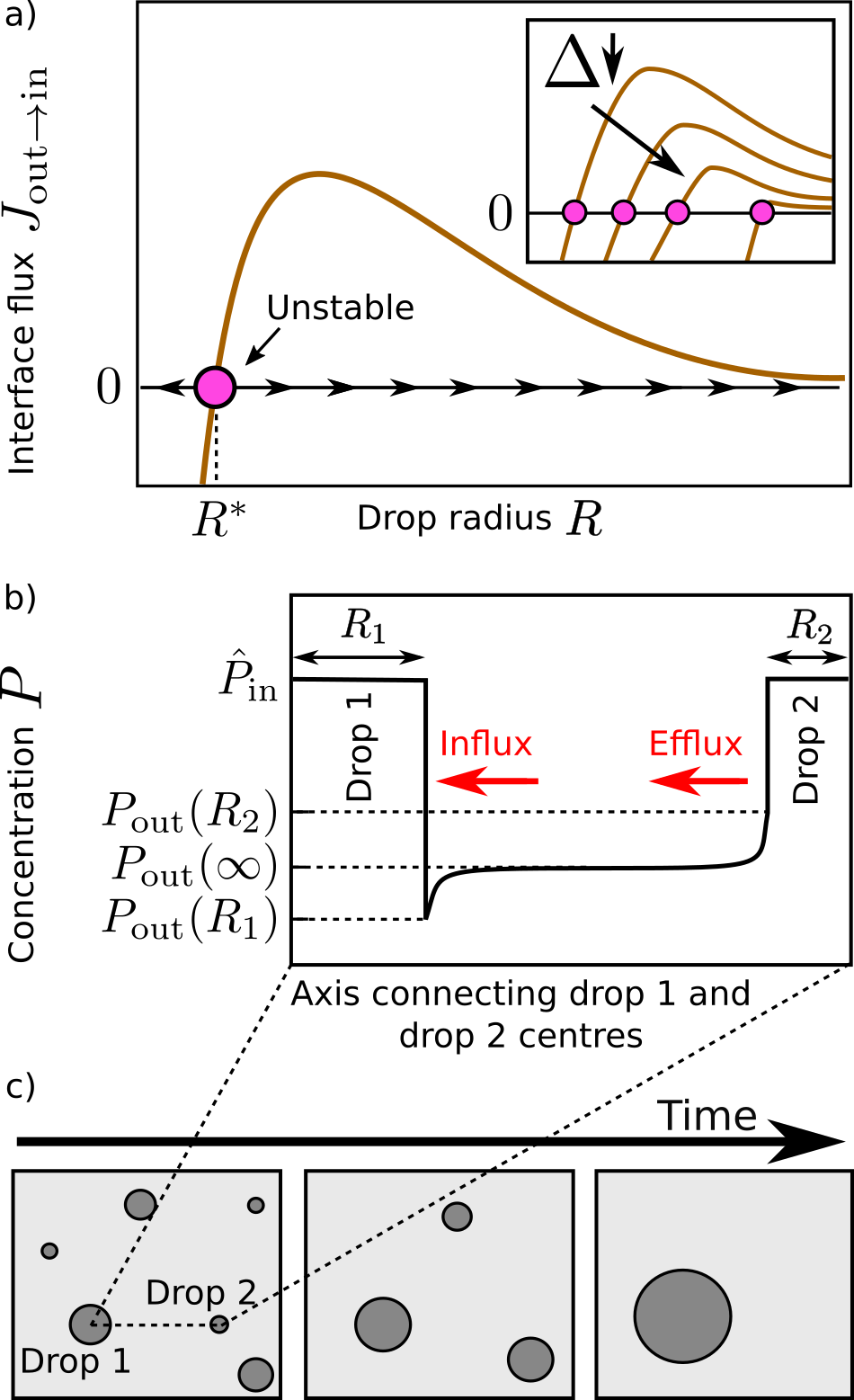}
	\caption
	{
		\textit{Ostwald ripening in equilibrium systems.}
		a) The influx $\Joi$ of molecules $P$ at a drop interface for varying drop radius $R$ and constant supersaturation $\Delta$ (\eq \eqref{Jeq}). The steady state $R^*$ ($J=0$, purple disk) is unstable: drops larger than $R^*$ grow while smaller ones shrink. Insert: as $\Delta$ decreases, the critical radius $R^*$ increases.
		b) Schematic of $P$ concentration profiles along an axis connecting the centres of two drops of different radii $R_1>R_2$. The Gibbs-Thomson relation (\eq \eqref{gtP}) dictates that the solute concentration is lower close to the large drop ($\Po(R_1)$) than close to the small drop ($\Po(R_2)$). This causes a diffusive flux from small drops to large drops (red arrows). 
		c) A multi-drop system is therefore unstable against Ostwald ripening: small drops evaporate and large drops grow. As fewer drops survive the supersaturation $\Delta$ decreases causing more drops to dissolve (see insert in a)). Eventually a unique drop remains in a finite system.
	}
	\label{fig:equi2}
	\label{fig:equi3}
\end{figure}

The diffusive profile leads to a flux $\Joi=D \nabla \Po |_{r=R}$ of molecules $P$ entering the drop at the interface \cite{lifshitz_jpcs61}:
\beqn
\label{Jeq}
\Joi &=&  \frac{D}{R} \left( \Delta - \frac{ \hatPo l_c }{R} \right) \ .
\eeqn
When the flux $\Joi$ is positive molecules $P$ accumulate at the interface leading to drop growth, while the drop shrinks when $\Joi<0$. We show in \fig \ref{fig:equi2} a) the flux $\Joi$ for varying drop radius $R$, assuming fixed supersaturation $\Delta$. There exists a steady state radius $R^*$ ($\Joi=0$, purple disk) that is unstable, called nucleus radius: smaller drops evaporate ($\Joi<0$, left arrows) and larger drops grow ($\Joi>0$, right arrows). This is due to the Gibbs-Thomson relation (\eq \eqref{gtP}) dictating that the concentration near a small drop is larger than that near a large drop. As a result a diffusive flux is directed from small to large drops (\fig \ref{fig:equi2} b), red arrows). A multi-drop system is therefore unstable against Ostwald ripening, i.e large drops grow at the expense of small ones (\fig \ref{fig:equi2} c)). As drops disappear, the average drop radius increases and the $P$ concentration near drops decreases according to the Gibbs-Thomson relation (\eq \eqref{gtP}). Hence the supersaturation $\Delta$ decreases because the total number of molecules $P$ in the system is fixed. This causes the critical radius $R^*$ to increase, as shown in the insert of figure \fig \ref{fig:equi2} a). Therefore Ostwald ripening occurs until, in a finite system, a single drop survive (\fig \ref{fig:equi2} c)) \cite{lifshitz_jpcs61}.


In the cell cytoplasm Ostwald ripening is not a desirable feature since it forbids the stability of multiple membrane-less organelles. In \Sec \ref{sec:noneq} we show how non-equilibrium chemical reactions in our ternary mixture can arrest Ostwald ripening.

\subsection{ Phase separation in presence of non-equilibrium chemical reactions}
\label{sec:noneq}

\begin{center}
	\begin{figure}[]
		\includegraphics[scale=0.95]{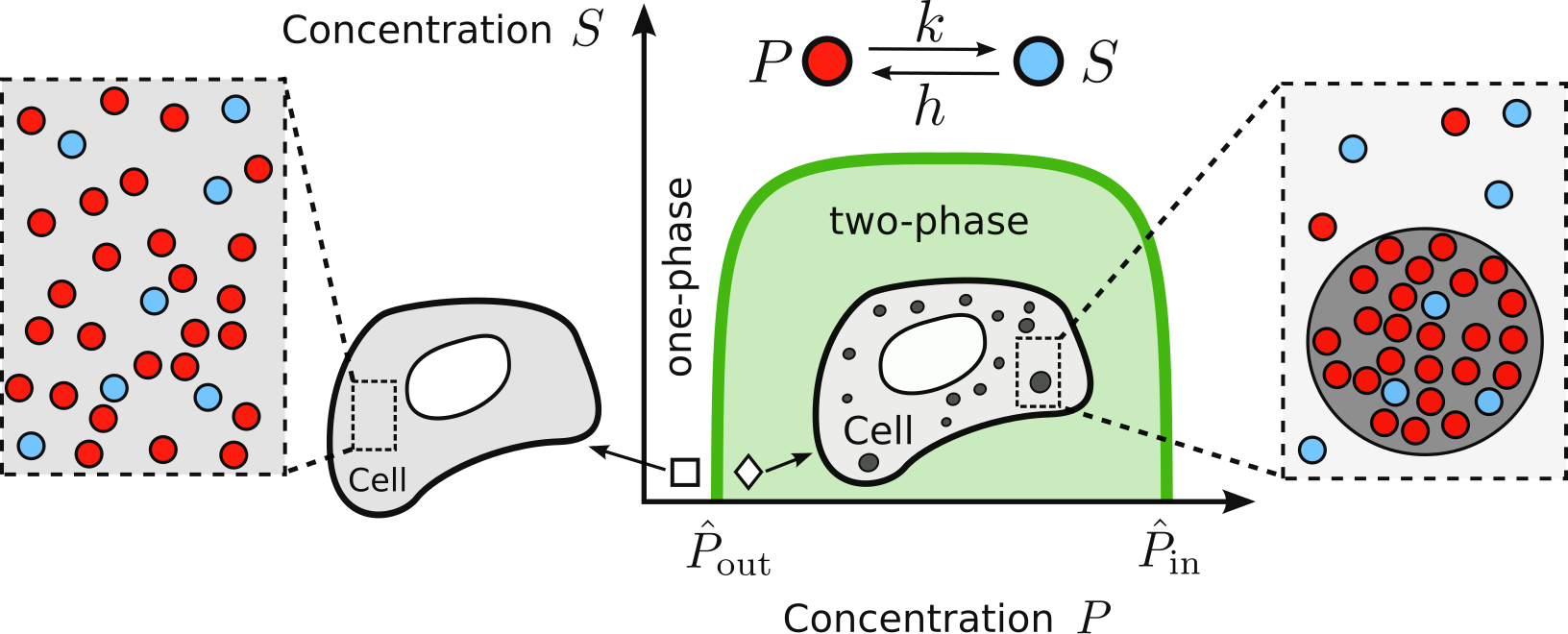}
		\caption{
			\textit{Model of non-equilibrium phase separation in the cell cytoplasm.} We consider the ternary mixture from \Sec \ref{sec:equi} with the addition of chemical reactions. Molecules $P$ can convert into $S$ with the rate constant $k$ and \textit{vice versa} with the rate constant $h$ (\eq \eqref{eq:reac}). $S$ does not participate to phase separation i.e. its concentration remains continuous at the drop interfaces ($\hat S_{\rm in} = \hat S_{\rm out}$).
			Figure reprinted from [Wurtz J.D. and Lee C.F., New Journal of Physics, vol. 20, no. 4, 045008, 2018], licensed under CC BY 3.0.
			\label{fig:noneq_intro}
		}
	\end{figure}
\end{center}

The presence of non-equilibrium chemical reactions have been proposed recently to explain multi-drop stability in the cytoplasm, as well as being a mechanism to control the formation, dissolution and size of membrane-less organelles \cite{zwicker_pnas14, zwicker_pre15, wurtz_prl18, wurtz_njp18}. We investigate in this section the physical mechanisms involved and recover the results from \cite{wurtz_prl18} by using a different and more intuitive approach.

\begin{figure}[]
	\centering
	\includegraphics[scale=1]{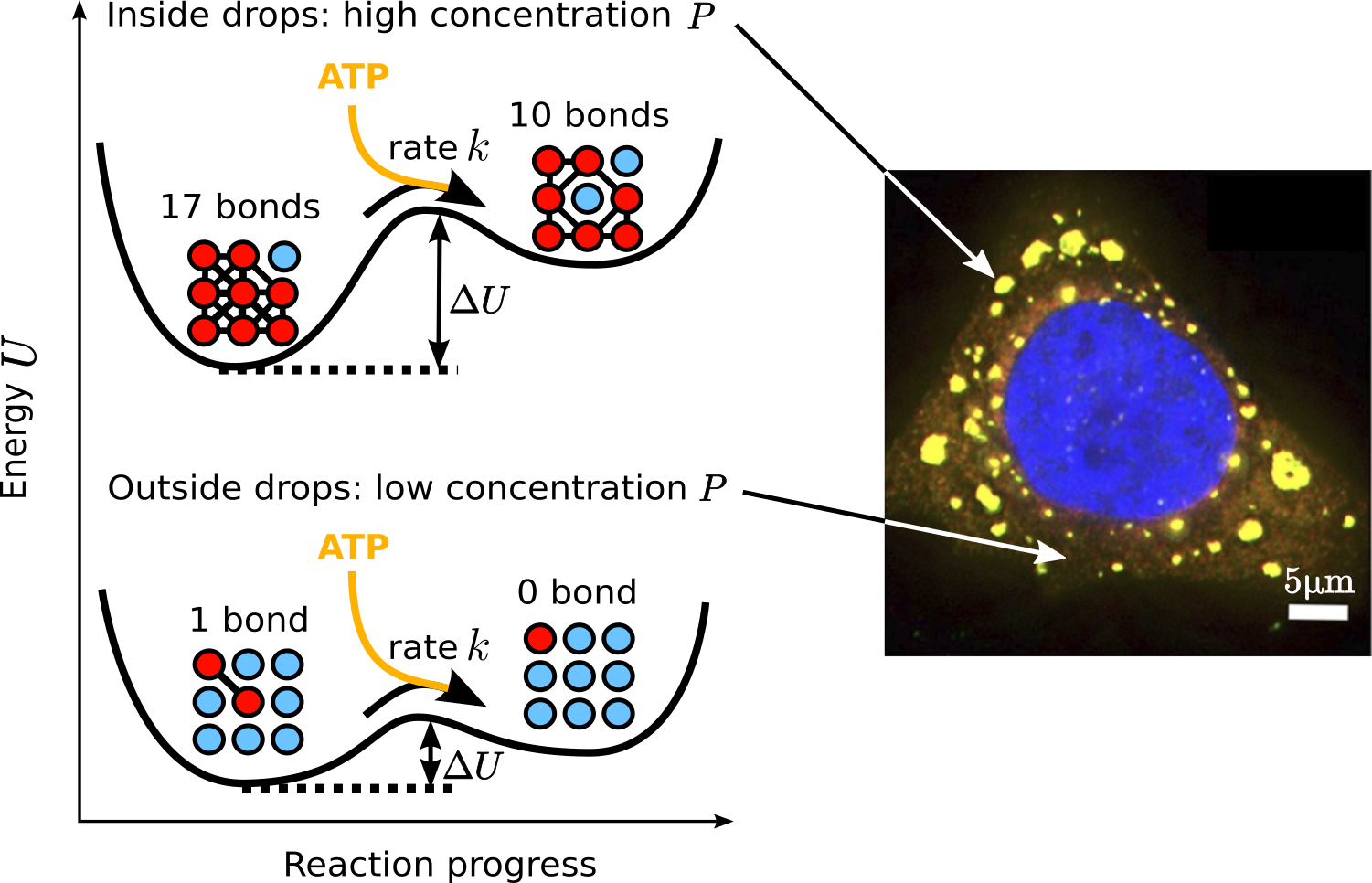}
	\caption
	{	
		\textit{Non-driven and driven chemical reactions in phase separating systems.}
		We consider here for clarity a particular example from our ternary mixture, where $P$ molecules (red disks) form bonds (black lines) with neighbouring $P$ molecules while molecules $S$ (blue disks) do not interact. We also concentrate on the forward reaction $P \rightarrow S$. The chemical conversion of a given $P$ into a $S$ requires an activation energy $\Delta U$ to break these bounds. Since drops are enriched in $P$, many bonds  need to be broken and $\Delta U$ is high (upper graph). Conversely the cytoplasm is poor in $P$ and $\Delta U$ is thereby small (lower graph). At thermal equilibrium the chemical conversions are non-driven and the energy required to overcome the barrier $\Delta U$ is provided by thermal fluctuations alone, and therefore the reaction rate constant $k$ decreases exponentially with $\Delta U$ \cite{hanggi_revModPhys90}. As a result, $k$ is thus concentration-dependent: $k$ is small inside drops where $\Delta U$ is high, and large in the cytoplasm. In the case of driven chemical reactions, such as ATP-dependent phosphorylation, an external source of energy is provided \cite{alberts_2008} (yellow arrows). Hence $k$ depends on the specificities of the reaction and can potentially be independent of $\Delta U$ i.e. concentration-independent. The rightmost part of the figure is adapted from [J.R. Wheeler et al., eLife vol. 5, e18413, 2018], licensed under CC BY 4.0.
	}
	\label{fig:energy}
\end{figure}

We consider the ternary mixture discussed in \Sec \ref{sec:equi} and now assume that $P$ and $S$ can be inter-converted by the chemical reactions:
\beqn
\label{eq:reac}
\ce{
	$P$
	<=>[$k$][$h$] 
	$S$ \ ,
}
\eeqn
where $k$, $h$ denote the forward and backward reaction rate constants (Fig.~\ref{fig:noneq_intro}). 
{\cblack In an equilibrium system $k, h$ should depend on the local concentrations. Indeed, since the mixture phase separates, the interactions between molecules $P$, $S$ or $C$ must be distinct, leading to concentration-dependent reaction rate constants, as detailed in \fig \ref{fig:energy}. In our non-equilibrium system however, we assume that energy consuming processes such as ATP-dependent phosphorylations drive the chemical reactions \cite{hill_book86} such that the rate constants $k$ and $h$ are concentration-independent.}


\begin{figure}
	\centering
	\includegraphics[scale=1]{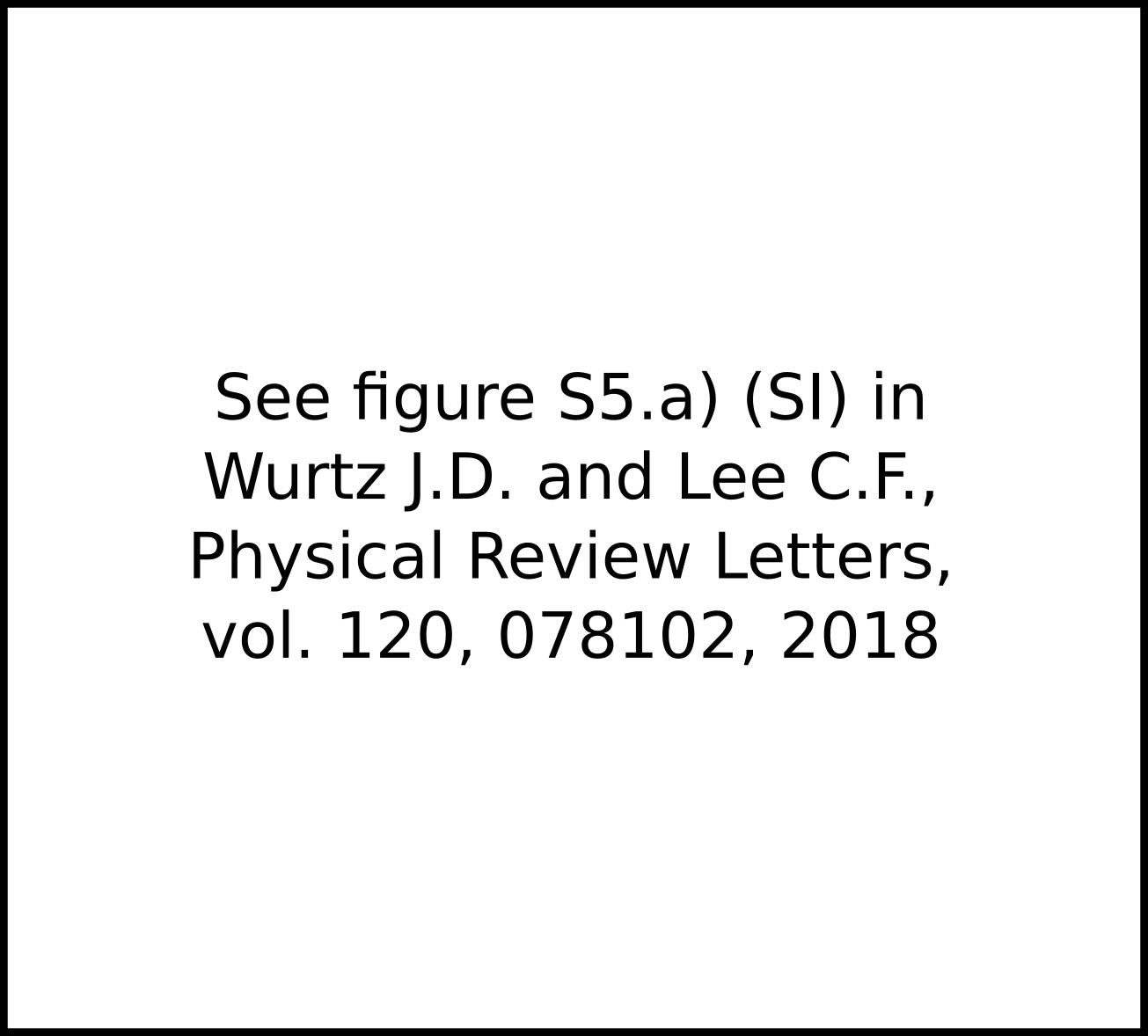}
	\caption{
		\textit{Monte Carlo simulation of phase separation in a ternary mixture with non-equilibrium chemical reactions} \cite{wurtz_prl18}. Molecules $P$ (red dots) form energetic bounds with neighbouring $P$, while molecules $S$ (blue dots) and $C$ (not shown) do not form bounds. Chemical reactions transform $P$ into $S$ and \textit{vice versa} with fixed reaction rate constants (\eq \eqref{eq:reac}). In the initial state molecules are randomly distributed (leftmost snapshot). As time progresses drops form and undergo Ostwald ripening leading to the increase of the average drops radius (middle snapshot). Eventually a multi-drop steady state is reached (rightmost snapshot). At the steady state drops have roughly the same radius and are evenly distributed.  Simulation details can be found in \cite{wurtz_prl18}. 
		\label{fig:MC} 
	}
\end{figure}

In \fig \ref{fig:MC} we show the results of Monte Carlo simulations of our non-equilibrium ternary mixture on a two-dimensional  lattice \cite{wurtz_prl18}. 
Molecules $P$ are represented by red dots, molecules $S$ by blue dots and molecules $C$ are not shown. Molecules $P$ form energetic bounds with neighbouring molecules $P$ to induce phase separation, and $P$ and $S$ randomly convert into each other with fixed transition probabilities according to \eq \eqref{eq:reac}. In the initial state all three types of molecules are distributed homogeneously (leftmost snapshot). Phase separation spontaneously occurs, $P$-rich drops appear and grow (middle snapshot), and the system undergoes Ostwald ripening leading to an increase of the average drop radius. Eventually a steady state is reached where the multi-drop system is stable and drops have similar radii (rightmost snapshot). Therefore the presence of the non-equilibrium chemical reactions arrest Ostwald ripening in our system. Interestingly drops tend to be of almost equal distance from each other, thus forming a close-packing lattice. In the next sections we will provide a theoretical analysis of the observed behaviour.
\\
\subsubsection{ Introduction of  a new length scale ($\xi$).}

We assume that our system remains close to equilibrium to the extent that local equilibrium applies. The interfacial region therefore remains governed by equilibrium principles and the coexistence concentrations at the interface $\Pio(R)$ are given by the Gibbs-Thomson relations (\GT). For simplicity we will assume that $S$ is inert to phase separation in the sense that its concentration inside and outside drops are identical ($\Sin(R)=\So(R)$). The more general scenario where $S$ can segregate inside or outside drops is treated in Ref. \cite{wurtz_prl18}. Since only $P$ phase separates we will refer to $P$ as the solute. {\cblack For simplicity} we assume the same diffusion coefficient $D$ for both species $P$ and $S$, and moreover that $D$ and the reaction rate constants $k$ and $h$ are identical both inside and outside drops.
In the context of membrane-less organelles this assumption is justified by the fact that drops are not highly packed but porous, and components rapidly shuttle in and out \cite{kedersha_jcb00,souquere_jcs09, bley_17}. The concentrations $P$ and $S$ therefore obey the following reaction-diffusion equations:
\beqn
\label{reacdiffP}
D \nabla^2 \Pio -k \Pio + h \Sio &=&0 \\
\label{reacdiffS}
D \nabla^2 \Sio + k \Sio  - h \Sio  &=&0 \ ,
\eeqn
where we used again the quasi-static assumption ($\pp_t P=0,\pp_t S=0$). In a homogeneous system $\nabla^2 P = \nabla^2S = 0$ and the concentrations are at their chemical equilibrium: $\Sio=\Pio k/h$.
Adding \eq \eqref{reacdiffP} and \eqref{reacdiffS} and imposing no-flux boundary conditions in drop centres and at the system boundary \cite{wurtz_prl18} we find that the combined concentration $P+S$ is homogeneous inside and outside drops:
\beqn
\label{P+S}
P_{\rm in,out}(r) + S_{\rm in,out}(r) = \phi_{\rm in,out} \ ,
\eeqn
where $\phi_{\rm in,out}$ are independent of $r$. 
The reaction-diffusion equations (\eqs \eqref{reacdiffP} and \eqref{reacdiffS}) therefore decouple and we obtain the solute concentration profiles:
\beqn
\label{profileFull}
\Pio(r) &=& U_{\rm in,out}^{(0)} + \frac{1}{r} \left( U_{\rm in, out}^{(1)} \ee^{r/\xi} + U_{\rm in, out}^{(-1)} \ee^{-r/\xi}\right) \ ,
\eeqn
with $U_{\rm in, out}^{(n)}$ independent of $r$ and
\beqn
\label{xi}
\xi = \sqrt{\frac{D}{k+h}}
\eeqn
is a new length scale introduced in the system by the chemical reactions, which is the length scale of the reaction-induced concentration gradients.


\begin{figure}[]
	\centering
	\includegraphics[scale=1.0]{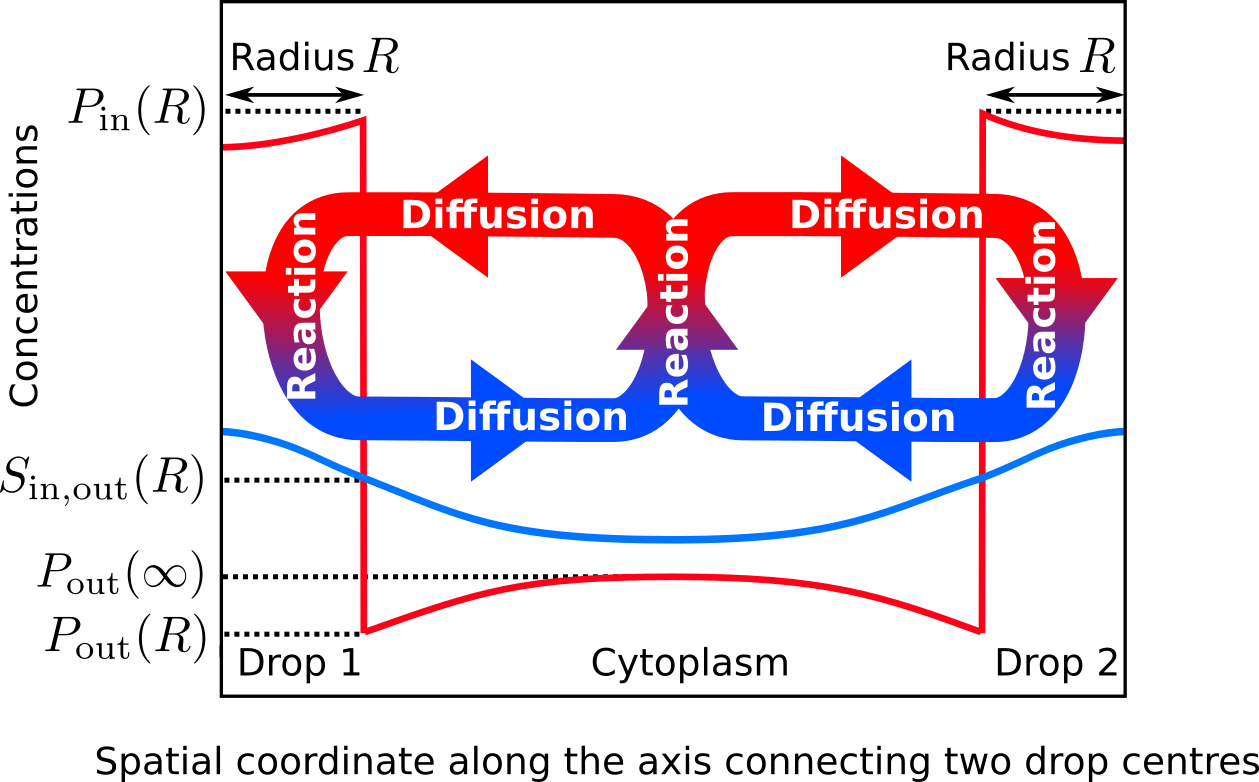}
	\caption
	{
		\textit{Non-equilibrium concentration profiles in a multi-drop system at steady state.} The profiles of $P$ (red curve) and $S$ (blue curve) (\eq \eqref{profileFull}) are shown along an axis connecting the centres of two drops of radius $R$. The interfaces are at local thermal equilibrium so the coexistence solute concentrations ($P_{\rm in,out}(R)$) are given by the Gibbs-Thomson relations (\eqs \eqref{gtIn} and \eqref{gtP}). The concentration $S$ is continuous at the interface ($\Sin(R)=\So(R)$). The non-equilibrium chemical reactions coupled with diffusion and phase separation create a circulating flux of molecules $P$ and $S$ between drops and cytoplasm. Drops are rich in $P$ so the chemical conversion $P \rightarrow_k S$ dominates, leading to an accumulation of $S$ molecules inside drops (downward red/blue arrows). The excess of $S$ is then transported by diffusion toward the cytoplasm (blue arrows). In the cytoplasm the reverse reaction dominates, leading to creation and accumulation of $P$ molecules between drops (upward blue/red arrow), which diffuse toward the drops (red arrows).
		\label{fig:profiles}
	}
\end{figure}

We show in \fig \ref{fig:profiles} a schematic of the concentration profiles of $P$ (red curve) and $S$ (blue curve) in a multi-drop system at steady state, along an axis connecting two drop centres. The chemically-induced concentration gradients and chemical reactions create a circulating flux of molecules $P$ and $S$ from drops to cytoplasm and \textit{vice versa}. Inside drops, the concentration of $P$ is high and the chemical conversion $P \rightarrow S$ dominates. Therefore the concentration of $S$ increases inside drops ({\cblack downward} red$\rightarrow$blue arrows). Molecules $S$ are then transported outside drops by diffusive fluxes (blue arrows). In the cytoplasm on the contrary, the reverse reaction $S \rightarrow P$ dominates so the concentration of $P$ increases between drops ({\cblack upward} blue$\rightarrow$red arrow). Molecules $P$ are then transported by diffusive fluxes toward drops (red arrows). 

We will now study how these fluxes can arrest Ostwald ripening, in two limiting regimes based on the drop size relative to the gradient length scale $\xi$ (\eq \eqref{xi}).

\subsubsection{ Small drop regime ($R \ll \xi$).}
\label{sec:small}

We start by considering the regime of small drop radius $R$ compared to the gradient length scale $\xi$ (\eq \eqref{xi}).
In a homogeneous system (not phase-separated) the concentrations are at their chemical equilibrium ($\nabla^2P=\nabla^2S=0$ in \eqs \eqref{reacdiffP} and \eqref{reacdiffS}). If the initial supersaturation is small, the cytoplasmic concentrations change only slightly during the phase separation and thus remains close to chemical equilibrium. We will therefore neglect chemical reactions in the cytoplasm and assume purely diffusive profiles outside the drops (\eq \eqref{profileEq}). The influx $\Joi$ of $P$ molecules at a drop interface is therefore identical to that in the equilibrium case (i.e. without chemical reactions, \eq \eqref{Jeq}). A quantitative analysis supporting this approximation is given in \ref{sec:app:cyto}.

\begin{figure}
	\centering
	\includegraphics[scale=1]{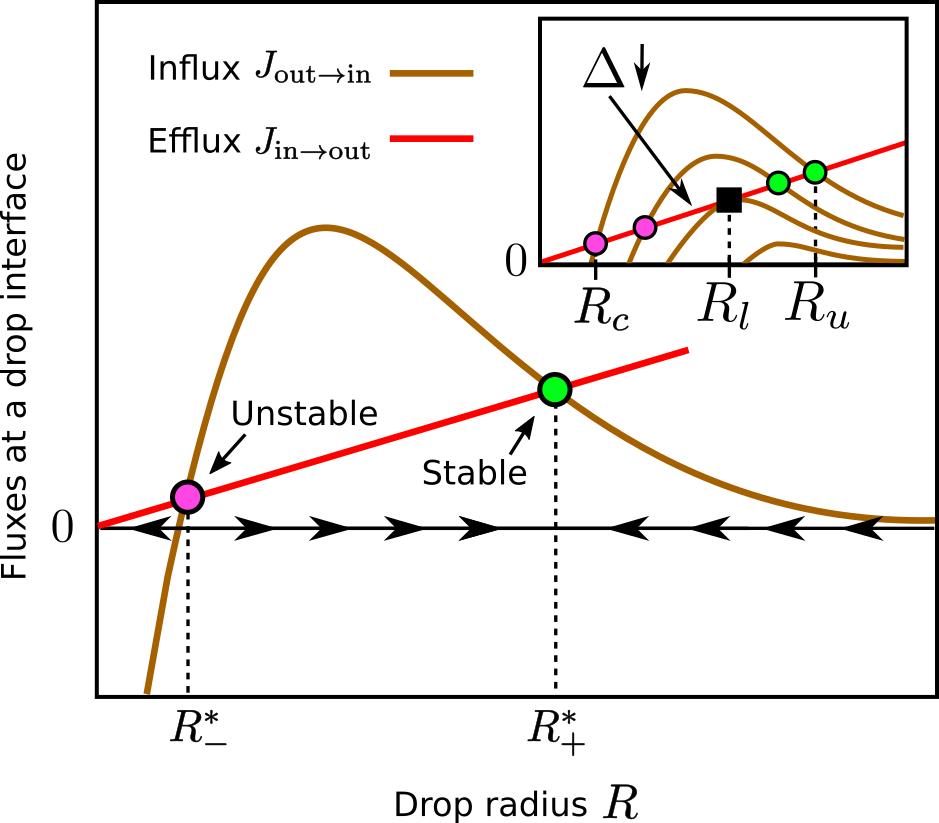}
	\caption{
		\textit{Molecular fluxes at a drop interface} for varying drop radius $R$ and fixed supersaturation $\Delta$. The efflux $\Jio$ (red curve, \eq \eqref{Jio_small}) driven by chemical reactions competes with the influx (brown curve) which is identical to that at the equilibrium condition (i.e without chemical reactions, see \eq \eqref{Jeq} and \fig \ref{fig:equi2} a)). Drops grow if $\Joi>\Jio$ and shrink otherwise. Two steady states radii $R^*_-<R^*_+$ exist. $R^*_-$ (purple circle), reminiscent from the equilibrium case, is unstable against Ostwald ripening: smaller drops shrink (left arrow) and larger drops grow (right arrows). At large radius the efflux $\Jio$ dominates so drops shrink. There exists therefore a second steady state $R^*_+$ (green disk) that is stable against Ostwald ripening: larger drops shrink and smaller drops grow. The insert shows the effect of decreasing the supersaturation $\Delta$. The smallest unstable steady state is labelled $R_c$, the largest stable steady state is labelled $R_u$, and the critical radius $R_l$ (black square) is the boundary between the unstable and the stable region (\eq \eqref{RcRlRu}).
	}
	\label{fig:noneq_small1}
\end{figure}

Inside drops, since $R\ll \xi$, the exponential terms in \eq \eqref{profileFull} are close to one. Imposing the no-flux boundary condition in the drop centre and 
enforcing the Gibbs-Thomson relation at the interface \eq \eqref{gtIn}  we get:
\beqn
\label{Pin_small}
\Pin(r)=\hatPin \ .
\eeqn
We find that the solute profile remains flat inside drops, unaffected by the chemical reactions. This can be interpreted as follow. The dominant reaction inside drops, $P\rightarrow_k S$, creates an excess of $S$ molecules. Since drops are small compared to the gradient length scale $\xi$ (\eq \eqref{xi}) the diffusion coefficient $D$ is large enough so that the excess of $S$ is quickly evacuated outside drops by diffusion, leaving the drop concentrations unperturbed. The subsequent efflux $J_{\rm in \rightarrow out}$ of molecules $S$ at the drop interface is therefore simply proportional to the degradation rate of $P$ molecules ($k \hatPin$) multiplied by the drop volume and divided by the drop surface:
\beqn
\label{Jio_small}
J_{\rm in \rightarrow out} &=& \frac{k \hatPin R}{3} \ .
\eeqn

We show in \fig \ref{fig:noneq_small1} the fluxes $\Jio$ (brown curve) and $\Joi$ (red curve) at the drop interface, for varying drop radius $R$ and at fixed supersaturation $\Delta$. The only difference with the equilibrium case (without chemical reactions, \fig \ref{fig:equi2} a)) is the existence of the chemical reaction-driven efflux $\Jio$, which competes with the influx $\Joi$. At small drop radius $\Joi$ is negligible, and the steady state $R^*_-$ (purple disk) is reminiscent from the equilibrium case  (\fig \ref{fig:equi2}): the system is unstable against Ostwald ripening, i.e. smaller drops dissolve ($\Jio > \Joi$) while larger drops grow ($\Jio < \Joi$). At large radius however the efflux $\Jio$ dominates causing drop shrinkage. This introduces a new steady state $R^*_+$ (green disk) that is stable against Ostwald ripening: larger drops shrink and smaller ones grow.

As $\Delta$ gets smaller, $R^*_-$ increases while $R^*_+$ decreases, as shown in the insert of \fig \ref{fig:noneq_small1}, and we can graphically identify three critical radii for which we provide approximate expressions \cite{wurtz_prl18}:
\beq
\label{RcRlRu}
R_c = \frac{\hatPo l_c}{\frac{ h \phi}{k+h}-\hatPo} \sep  R_l = \left( \frac{3Dl_c \hatPo}{2 k \hatPin} \right)^\frac{1}3 \sep R_u = \sqrt{\frac{D \left( \frac{h \phi}{k+h} - \hatPo\right)}{k \hatPin}}  \ .
\eeq
where $\phi$ is the average concentration of $P+S$ in the entire system. {\cblack Due to the incompressibility condition the overall solvent concentration $C$ in the entire system is given by $\psi - \phi$, where we recall that $\psi$ is the combined concentration of all species $P+S+C$ (\Sec \ref{sec:equi}).}
$R_c$ is the smallest unstable steady state $R^*_-$, $R_u$ the largest stable steady state $R^*_+$, and $R_l$ is the boundary between the unstable and the stable regimes (black square symbol).  We note that similar scaling laws to $R_l$ and $R_u$ have been previously found in binary mixtures i.e. without the solvent $C$ \cite{zwicker_pre15}.
There exists also a critical forward rate constant $k_c$ above which  the slope of $\Jio$ is so large that all drops dissolve ($\Jio>\Joi$ for all $R$). $k_c$ is maximally bounded as follow \cite{wurtz_prl18}:
\beqn
\label{kc}
k_c < {\rm min} \left[\frac{\phi-\hatPo}{\hatPo} h\ ;~ \frac{4 D \left(\phi-\hatPo\right)^3}{9 l_c^2\hatPin\hatPo^2} \right]
\ . 
\eeqn
When $k>(\phi-\hatPo)/\hatPo h$ the conversion $P\rightarrow_k S$ is so fast that the system is outside the phase-separating region (``$\diamondsuit$" symbol in \fig \ref{fig:noneq_intro}).

\begin{figure}
	\centering
	\includegraphics[scale=0.9]{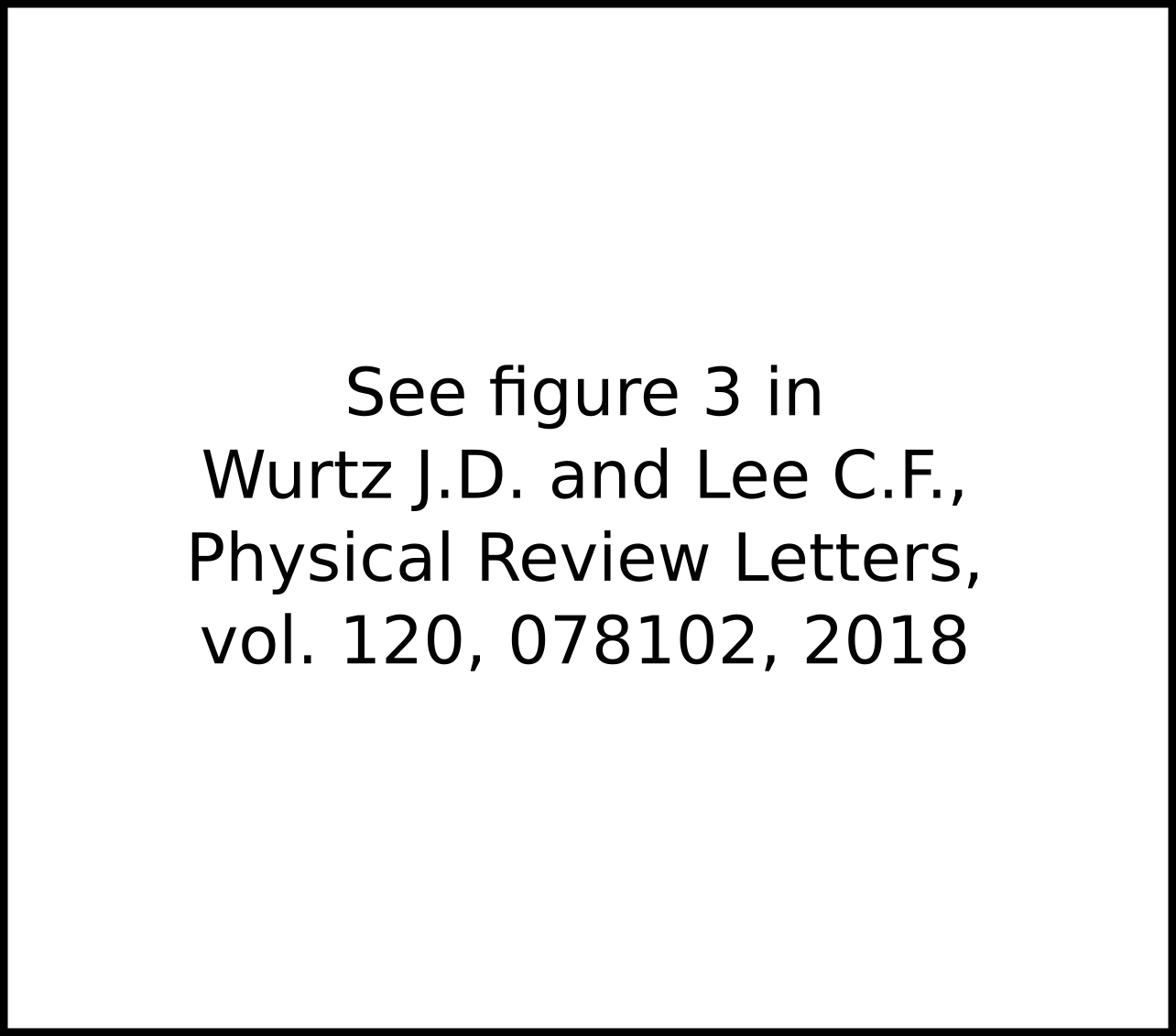}
	\caption{
		Stability diagram of a multi-drop system  for varying forward rate constant $k$ and drop radius $R$. A steady state exists within the continuous curve. Outside this region all drops dissolve (downward arrows). Outside the grey region but within the continuous curve, drops are unstable against Ostwald ripening and coarsen (upward arrows). In the grey region drops are stable against Ostwald ripening. The stability-instability boundary is shown by a dashed curve. The analytical results for the critical radii in the small drop regime are showed by the upper dotted line ($R_u$,  \eq \eqref{RcRlRu}) and lower dotted line ($R_l$). The analytical results for the critical rates $k_c$, $k_l$ and $k_u$ (\eqs \eqref{kc}, \eqref{k_l} and \eqref{k_u}) are showed by arrows. Large drops dissolve beyond $k_u$, all drops dissolve beyond $k_c$, and large drops are stable against Ostwald ripening beyond $k_l$. 
		Parameters: $h=10^{-2} s^{-1},~ l_c=10^{-2}{\rm \mu m},~ D=1 {\rm \mu m^2 s^{-1}},~\hatPin=10^{-1}{ \nu^{-1}},~\hatPo=10^{-4}{\nu^{-1}}$, $\phi=5 \times 10^{-4} \nu^{-1}$, where $\nu$ is the molecular volume of $P$ and $S$ and can be chosen arbitrarily. 
		\label{fig:all}
	}  
\end{figure}

We show the stability diagram in \fig \ref{fig:all} (lower region, small drop regime) for varying forward rate constant $k$ and drop radius $R$. A multi-drop system at steady state $R$ is unstable against Ostwald ripening if $R_c<R<R_l$ (upward arrows), and stable if $R_l<R<R_u$ (grey region).

As the rate constant $k$ decreases the drop radii tend to increase ($R_u, R_l$). When the radii get comparable or larger than the gradient length scale $\xi$, the system exits the small drop regime and enters the large drop regime, which we will now discuss.

\subsubsection{Large drop regime.}
\label{sec:large}

\begin{figure}
	\centering
	\includegraphics[scale=1]{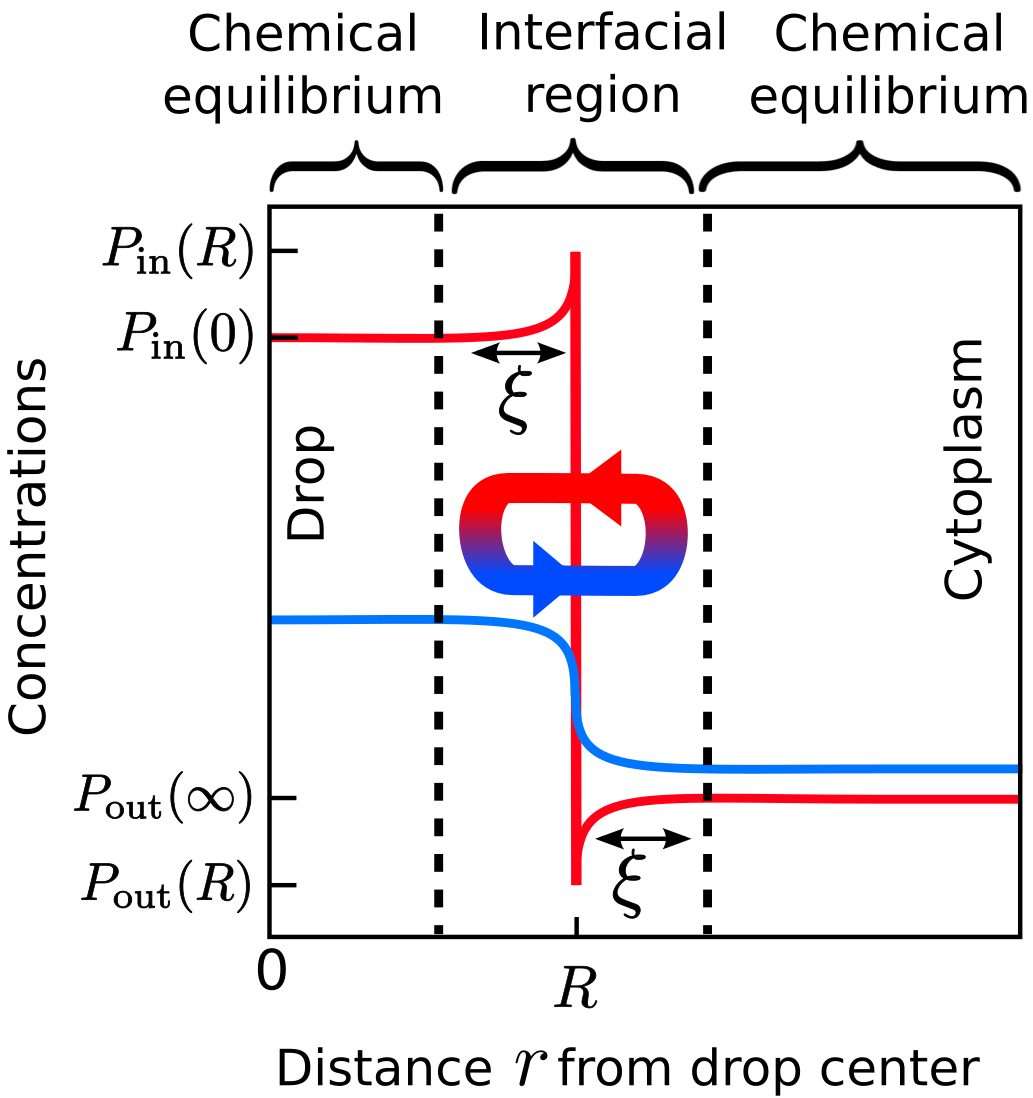}
	\caption{
		\textit{Concentration profiles in the large drop regime.}
		The profiles of $P$ (red curve) and $S$ (blue curve) are flat and therefore at chemical equilibrium far from the interface ($\Pin(0),\Pinf$). Concentration gradients of length scale $\xi$ (\eq \eqref{xi}) are localised at the interfacial region only. Hence the reaction-diffusion-driven circulation of molecules $P$ and $S$ between drops and cytoplasm (blue$\leftrightarrow$red circular arrow) is also localised at the interfacial region only. See \fig \ref{fig:profiles} for details about this circulation. 
		\label{fig:noneq_large1}
	}  
\end{figure}

We now concentrate on the regime where drops are large compared to the gradient length scale ($R \gg \xi$, \eq \eqref{xi}). The quantity $\xi/r$ in the profile expression \eq \eqref{profileFull} is therefore small in the cytoplasm ($r>R$). Then, enforcing the no-flux boundary condition in drop centre and the Gibbs-Thomson relation at the interface (\eqs \eqref{gtP}), the profiles become \cite{zwicker_pre15,wurtz_prl18}:
\beqn
\label{profileLargeIn}
\Pin(r) &=& \Pin(0) + \left( \Pin(R) - \Pin(0) \right) \frac{R}{r} \frac{\sinh(r/\xi)}{\sinh(R/\xi)} \\
\label{profileLargeOut}
\Po(r) &=& \Po(\infty) - \left( \Po(\infty) - \Po(R) \right) \frac{R}{r} \ee^{-(r-R)/\xi} \ ,
\eeqn
and $S_{\rm in,out}(r)=\phi_{\rm in,out} - P_{\rm in,out}(r)$, with $\phi_{\rm in}$ being independent of $r$ (\eq \eqref{P+S}). We show in \fig \ref{fig:noneq_large1} a schematic of the concentration profiles of $P$ and $S$ along an axis passing through a drop centre. Far from the interface the profiles are flat, implying that $P$ and $S$ are in chemical equilibrium ($\nabla^2 P = \nabla^2 S = 0$ in \eqs \eqref{reacdiffP} and \eqref{reacdiffS}). Concentration gradients and fluxes exist only in the interfacial region.

Interestingly this large drop regime cannot exist in the absence of the solvent ($C$), i.e. in a binary mixture. Indeed, imposing in this case the incompressibility condition $\Pio(r)+S_{\rm in,out}(r)={\psi}$  with $\psi$ being independent of $r$, and the chemical equilibrium condition $S_{\rm in,out}(r)=\Pio(r) k/h$, there is a unique solution for the solute concentration: the solute concentration, both inside and outside drops away from the interfaces, equals the overall solute concentration in the whole system. Since the solute concentration inside and outside drops is supersaturated (``$\diamondsuit$" symbol in \fig \ref{fig:noneq_intro}), new drops will be created through further phase separation,
ultimately leading  the system to the small drop regime (\Sec \ref{sec:small}). In our ternary mixture however the incompressibility condition is $P_{\rm in,out}(r)+S_{\rm in,out}(r)+C_{\rm in,out}=\psi$ with $C_{\rm in,out}$ independent of $r$ due to \eq \eqref{P+S}. Since $C$ can distribute differentially inside and outside drops ($C_{\rm in}\neq C_{\rm out}$), and adding now the chemical equilibrium conditions $S_{\rm in,out} = \Pio k/h$ we find two distinct solutions for the solute concentration. In other words the added degree of freedom from the solvent allows for distinct chemical equilibrium concentrations inside and outside drops, as shown in \fig \ref{fig:noneq_large1} ($\Pin(0), \Po(\infty)$).

Using the solute concentration profiles \eqs \eqref{profileLargeIn} and \eqref{profileLargeOut} we obtain the influx of solute $P$ ($\Joi = D \dd \Po/\dd r|_{R}$) and the efflux of molecules $S$ ($\Jio = -D \dd \Sin/\dd r|_{R}= D \dd \Pin/\dd r|_{R}$, \eq \eqref{P+S}):
\beqn
\label{Joi_large}
\Joi &=&   \frac{D  \left( \Pinf - \Po(R) \right)}{\xi}  \left( 1 + \frac{\xi}{ R} \right) \\
\label{Jio_large}
\Jio &=& \frac{D \left( \Pin(R) - \Pin(0) \right) }{\xi} \left( 1-\frac{\xi}{R} \right) \ ,
\eeqn
Neglecting for the time being the terms $\xi/R$ which are small in this regime, we find that both fluxes are simply proportional to the difference of concentrations between the interface and far from the interface. As the drop radius $R$ increases, the solute cytoplasmic concentration $\Po(R)$ close to the interface decreases according to the Gibbs-Thomson relation (\eq \eqref{gtP}). This causes an increase of the solute influx $\Joi$ into the drop, and therefore contributes to further drop growth. Hence, as in equilibrium systems (i.e without chemical reactions, \Sec \ref{sec:equi}), the Gibbs-Thomson relation supports Ostwald ripening.

Let us now study the effect of the term $\xi/R$ in the flux expressions \eqs \eqref{Joi_large} and \eqref{Jio_large}. The term $\xi/R$, which vanish at large radius $R$, captures the concentration profile asymmetry inside and outside drops (\eqs \eqref{profileLargeIn} and \eqref{profileLargeOut}) arising from the spherical drop shape. As a result drop expansion tends to increase the influx $\Joi$ while decreasing the efflux $\Jio$, acting against further expansion. The chemical reaction-induced term $\xi/R$ therefore tends to stabilise a multi-drop system against Ostwald ripening.


In summary in this large drop regime we found that the Gibbs-Thomson relations tend to destabilise a multi-drop system causing Ostwald ripening, and chemical reactions on the contrary have a stabilisation effect, as in the small drop regime (\Sec \ref{sec:small}). In \ref{sec:app:large} we provide a quantitative analysis of \eqs \eqref{Joi_large} and \eqref{Jio_large} and find the critical forward rate $k_l$ beyond which Ostwald ripening is arrested:
\beqn
\label{k_l}
k_l = \frac{2 l_c \hatPo}{D^\frac{1}{2} \hatPin} h^\frac{3}{2} \ ,
\eeqn
and the maximal forward rate constant $k_u$ above which drops dissolve:
\beqn
\label{k_u}
k_u = 2\frac{\phi-\hatPo}{\hatPin} h \ ,
\eeqn
and we have recovered the results from \cite{wurtz_prl18}.
Note that for $k>k_u$ drops may still exist in the small drop regime (\Sec \ref{sec:small}).
We show in the stability diagram in \fig \ref{fig:all} \cite{wurtz_prl18} the stable and unstable regions depending on the forward rate constant $k$ and drop radius $R$ 
We also note that there is no upper-bound on the drop radius $R$ contrary to the small drop regime (\Sec \ref{sec:small}).

\subsection{  Spatial organisation}
\label{sec:spatial}

\begin{figure}
	\centering
	\includegraphics[scale=0.9]{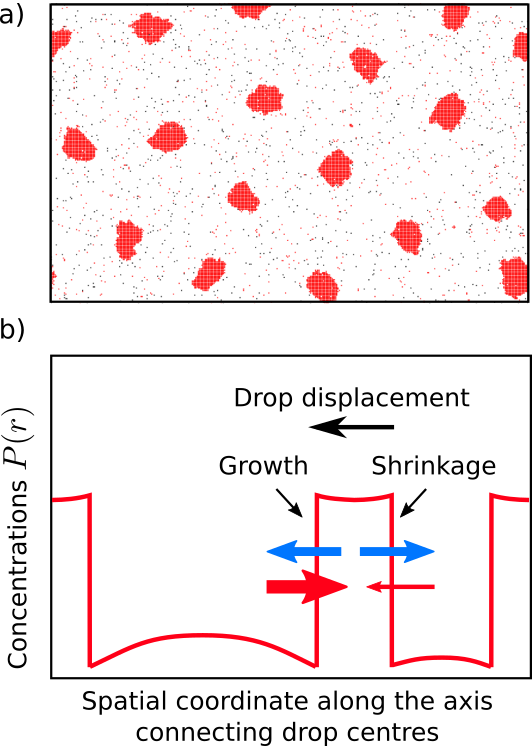}
	\caption{
		\textit{Drop spatial organisation.}
		a) Snapshot of Monte Carlo simulations of a multi-drop system at steady state (no momentum transfer between molecules, see main text for details). Drops spontaneously distribute homogeneously in space. b) Schematic of the solute profile across three adjacent drops (red curves). The influx of molecules $P$ is shown by red arrows and the efflux of molecules $S$ by blue arrows. The solute profile gradient between drops far from each other is stronger than between close drops. The solute influx is therefore larger on one side of the drop (thick red arrow) and weaker on the opposite side (thin red arrow), causing the drop to grow on one side and dissolve on the other side. This leads to effective drop displacement tending to align drops on a lattice.
		\label{fig:spatial}
	}  
\end{figure}

Another interesting phenomenon resulting from this type of non-equilibrium phase separation is the potential spontaneous spatial organisation of drops on a lattice, as observed in Monte Carlo simulations shown in \fig \ref{fig:spatial}a \cite{wurtz_prl18}. In this section we provide a simple intuitive argument that accounts for the observed lattice organisation. Note that these simulations do not capture momentum transfer between molecules, hence drop diffusion is not reproduced accurately. Whether this lattice structure survives in the presence of drop thermal diffusion remains to be investigated.

At steady state we have seen that a flux balance exists at the interface between the influx of solute $P$ and the efflux of $S$ (\fig \ref{fig:profiles}). 
We now consider a system where drops are regularly distributed on a lattice initially, and study what happens when a drop moves away from its initial position (\fig \ref{fig:spatial}b)). 
Since the reaction $S \rightarrow P$ dominates in the dilute phase between drops, this leads to an accumulation of molecules $P$ in the cytoplasm, and a subsequent diffusive flux of $P$ toward drops (\fig \ref{fig:profiles}). As the time required for $P$ molecules to reach drops increases with the inter-drop distance, $P$ accumulates predominantly where the inter-drop distance is large. A similar intuitive argument was also provided to explain the concentration profiles inside drops (\eq \eqref{Pin_small}). In addition, due to the concentration $P$ at a drop interface being fixed by the Gibbs-Thomson relation (\eq \eqref{gtP}), the concentration gradient of $P$ is increased on the side of the drop where the inter-drop distance is the largest. The influx of solute on this side of the drop is therefore strengthened (thick red arrow). On the contrary on the side of decreased inter-drop distance, a weaker solute gradient leads to a smaller solute influx (thin red arrow).  The imbalance of the solute influx on each side of the drop causes effective drop displacement toward the larger influx, i.e., toward the drop initial position.
As a result, chemical reactions in our multi-drop system tend to distribute drops on a lattice structure.

\section{Active matter: motile organisms in the incompressible limit}
Active matter refers to physical systems in which some or all constituents of the system can exert forces continuously on their surrounding environment \cite{marchetti_rmp13}. For instance, in the case of a bird flock, the birds fly by flapping their wings to move the air around them; in the case of a cell tissue on a substrate, the cells move  via coordinated and ATP-driven remodelling of biopolymers beneath their cell membranes \cite{salbreux_trends12}. Active matter constitutes a non-equilibrium system and the energy is provided either through a continuous supply of fuel or by energy already stored in the system. 

Here, we will focus exclusively on active matter in the condensed state, to the extent that the system can be viewed as incompressible. Biological examples include a dense collection of motile bacteria \cite{Wensink2012}, or a cell tissue in which the cells are undergoing dynamic rearrangement \cite{rossen_natcomm16,mortia_devcell17}. In the hydrodynamic limit (the limits of long time and long wavelength), an active matter can usually be described by  equations of motion ({\bf EOM}) of  field variables that correspond to some coarse-grained properties of the system, such as the local density  and the local velocity \cite{toner_annphys05, Julicher2007,marchetti_rmp13}. Such an  EOM can generically be written down based on symmetry consideration alone and the associated universal behaviour of the system can then be analysed using analyical methods such as dynamical renormalisation group ({\bf DRG}) methods \cite{Ma1975,forster_pra77}, or  numerical methods. In this review, we will focus on the former approach and discuss how it enables us to elucidate the universal behaviour of an incompressible active fluid at criticality and in the ordered phase in certain spatial dimensions.

\subsection{Hydrodynamic theory of incompressible passive fluids -- Navier-Stokes equation}
For an equilibrium system, symmetry constrains the allowable form of the Hamiltonian of the system \cite{Kardar2007}. For a  non-equilibrium system, although a Hamiltonian may no longer be relevant, we can still use symmetry to deduce the form of EOM \cite{toner_prl95,toner_pre98}. To illustrate this approach, we will now review how such a symmetry consideration can help us  derive the incompressible Navier-Stokes equation. 

In an incompressible fluid, the obvious field variable is the velocity field $\bv(\bbr,t)$, whose dynamics can be  written as:
\beq
\pp_t {\bv} = \frac{\bf F}{\rho}
\ ,
\eeq
where $\rho$ is the density field and ${\bf F}$ corresponds to the local force density. Since the system is incompressible, $\rho$ is constant everywhere and we will ignore this constant factor from now on.

We now impose the following symmetries:
\begin{enumerate}
	\item
	{\bf Temporal invariance}: ${\bf F}$ does not depend on time $t$ explicitly, hence forbidding terms like $t\bv$.
	This symmetry means that experimental results on the fluid motion do not depend on the day of the week on which  the experiments are done.
	\item
	{\bf Translational invariance}:  ${\bf F}$ does not depend on the spatial location ${\bf r}$ explicitly, hence forbidding terms like $\bbr$. This symmetry means that experimental results  do not depend on the location where the experiments are done
	\item
	{\bf Rotational invariance}: the EOM is invariant if the reference frame is rotated, hence forbidding terms like $\bw$ for some constant vector $\bw$  since a term like this will pick out a particular direction in the system. This symmetry means that experimental results do not depend on  which direction the experimental apparatus are positioned towards. 
	\item
	{\bf Parity invariance}: the EOM is invariant under spatial inversion, hence forbidding terms like $\nabla \times \bv$. This symmetry means that the physical system has no chirality, i.e., the physics of fluid motion has no handedness.
\end{enumerate}

Imposing these symmetries, and expanding ${\bf F}$ in powers of $\bv$ and of the spatial derivatives $\nabla$, we arrive at the generic EOM:
\beq
\label{AMeq:fulleom}
\pp_t \bv =-\vec{\kappa} -\lambda (\bv \cdot \nabla) \bv+\mu \nabla^2 \bv +(a-bv^2) \bv {\cblue + \mu' (\nabla^2)^2 \bv+cv^4 \bv+\ldots}
\eeq
where $v\equiv |\bv|$ and ``{\cblue $\ldots$}'' refer to higher order  terms permissible in ${\bf F}$ that are not shown.  Note that the first term on the R.H.S. of the above equation, $-\vec{\kappa}$, is a vectorial Lagrange multiplier there to enforce the incompressibility condition $\nabla \cdot \bv =0$. By the Helmholtz decomposition, we can write ${\bf F}$ as $\nabla p + \nabla \times {\bf A}$ where $p$ is a scalar field and ${\bf A}$ is a vector field  \cite{stone_b09}. Since we want to subtract off part of ${\bf F}$ that is {\it not} divergence-free, we have $\vec{\kappa} = \nabla p$.

Our  EOM so far does not look like the Navier-Stokes equation yet as we are still missing one crucial symmetry: the Galilean invariance.
\begin{enumerate}
	\setcounter{enumi}{4}
	\item	
	{\bf Galilean invariance}: when no external forces are acting on the system, the EOM is invariant if the reference frame is boosted to another reference frame that is travelling at a constant speed in an arbitrary direction.
\end{enumerate}
Under this additional symmetry, the EOM remains invariant if we perform the following simultaneous transformations: $\bbr \mapsto \bbr-\bw t$ and $\bv(\bbr,t) \mapsto \bv(\bbr-\bw t,t) +\bw$, for some arbitrary vector $\bw$. Imposing this constraint, the EOM, to order $\cO(\nabla^4)$, is
\beq
\pp_t \bv + (\bv \cdot \nabla) \bv=-\nabla p +\mu \nabla^2 \bv
\ ,
\label{AMeq:NS}
\eeq
which is exactly the incompressible Navier-Stokes equation, with $p$ interpreted as the pressure divided by the density.
If we are only interested in the coarse-grained (long wavelength) behaviour of the system, we can argue that higher order terms, such as $\nabla^4 \bv$,  are unimportant compared to $\nabla^2 \bv$, and thus \eq (\ref{AMeq:NS}) can be viewed as the hydrodynamic equation of incompressible fluids.

In a physical system, fluctuations, e.g., thermal in origin, are inevitable. Using the fluctuation-dissipation relation, fluctuations can be added to the above Navier-Stokes equation that renders it suitable to describe incompressible thermal fluids \cite{landau1980statistical}. Analytical treatment of the resulting stochastic partial differential equation can then be used to elucidate the {\it universal} behaviour of the system. For instance, the existence of long-time tail of various correlation functions of thermal fluids, first discovered via simulations \cite{alder_prl67,alder_pra70}, have been confirmed using diverse analytical methods such as kinetic theory \cite{dorfman_prl70,ernst_prl70} and DRG analysis \cite{forster_pra77}. We will not review these well known results here, instead we will now  turn our attention to incompressible active fluids.

\subsection{Incompressible active fluids}
We will focus exclusively on the so-called ``dry'' active matter \cite{toner_annphys05,marchetti_rmp13}, in the sense that there exists a fixed background in the system for the active constituents to exert forces on. Experimentally, the active constituents can be motile cells and the fixed
background can be a gel substrate that the cells crawl on. In contrast, wet active matter describes motile organisms in a fluid medium in which organisms move by exchanging momentum with the surrounding fluid, and the resulting fluid flow can in turn affect the motion  of the organisms \cite{Simha2002,Hatwalne2004}.  

In dry active matter, due to the ability of each active volume element to generate forces against a fixed background, the Galilean invariance no longer applies.  Omitting this symmetry, the general EOM of a generic incompressible active fluids is of the form of \eq (\ref{AMeq:fulleom}), which is in fact exactly the incompressible version of the Toner-Tu equation devised to describe the flocking behaviour \cite{toner_prl95,toner_pre98,toner_pre12}.

\begin{figure}
	\centering
	\includegraphics[scale=.5]{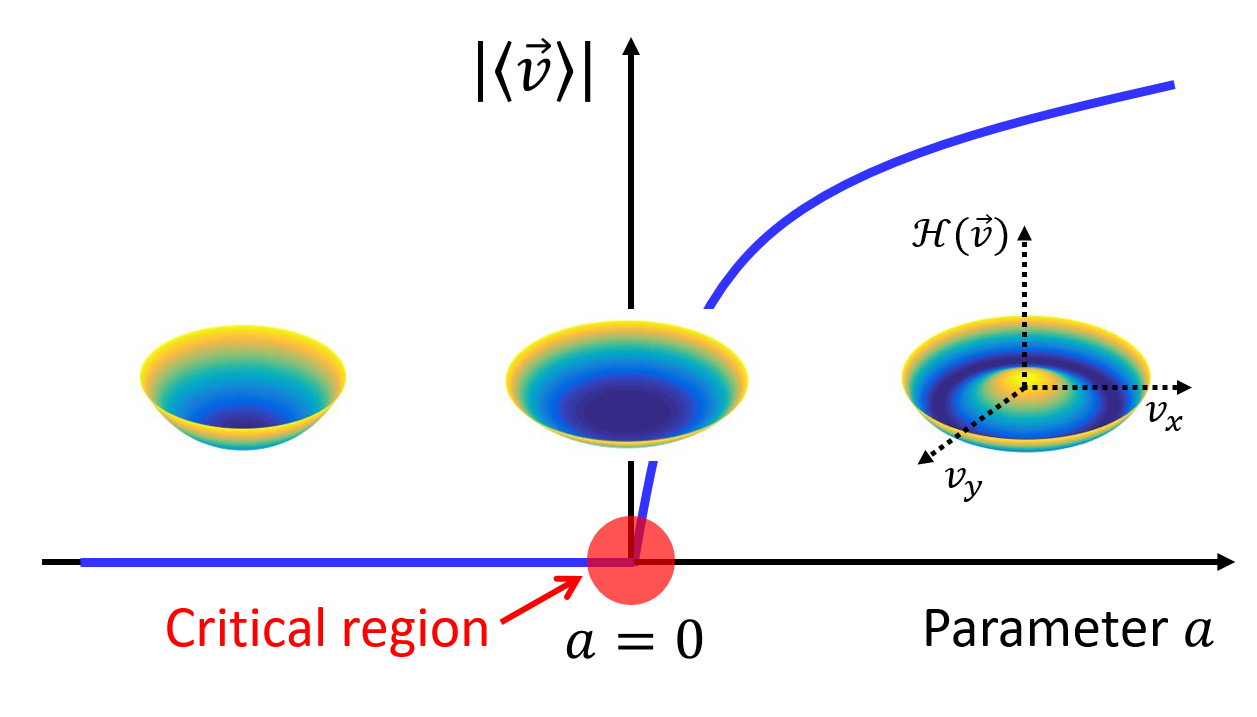}
	\caption{
		In a generic incompressible active fluid, two distinct phases are possible. At the mean-field level, a disordered phase exists when the parameter $a$ is negative, and an ordered phase (characterised by a non-zero mean speed of the system) emerges when $a$ is positive. The transition between these two phases is continuous, or critical (region depicted in red). The surface plots depicts the ``potential energy landscape'' at the mean-field level for an active fluid in two dimensions. In the disordered phase, the energy landscape is like a parabolic bowl, while at the transition, the global minimum of the landscape becomes very flat. The landscape transitions further into the shape of a Mexican hat in the ordered phase.
		\label{fig:AMphase}
	}  
\end{figure}

Ignoring the blue terms in \eq (\ref{AMeq:fulleom}) for the time being (whose omissions will be justified later), and focusing on spatially homogeneous states (so all terms involving $\nabla$ become  zero), the simplified EOM can be written as
\beq
\pp_t \bv =-\frac{\delta \cH}{\delta \bv}
\eeq
where $\cH(\bv) = -a \bv^2/2 +b\bv^4/4$. $\cH$ can be viewed as a ``potential energy'' term, whose forms, depending on the parameter $a$, are depicted in \fig \ref{fig:AMphase} for a two dimensional system. When $a$ is negative, $\cH$ has only one minimum at $\bv=0$, which suggests that the only steady-state solution is the $\bv=0$ homogeneous state. We call this the {\it disordered} phase. As $a$ increases beyond zero, a continuum of minima emerges and all of these will have a non-zero mean speed. This corresponds to the {\it ordered} phase, or the {\it collective motion} phase. The transition between these two phases is continuous and thus constitutes a {\it critical transition}. 
From equilibrium statistical mechanics, we know that when spatial heterogeneity and fluctuations are restored,  the system	
can possess scale-invariant features at criticality \cite{Cardy1996,Goldenfeld1992,Kardar2007}. This is also what happens in our active fluid system, as we shall show next.

\subsection{Universal  behaviour at the critical point}
To understand the emergence of scale-invariant structures at the critical point when the system transitions from the disordered phase to the ordered phase, we will first analyse the EOM at the linear level and then incorporate the nonlinear effects using DRG methods.

\subsubsection{Linear theory.}
To arrive at the linear equation, we tune all the coefficients in the EOM to zero except for the terms below:
\beq
\pp_t \bv = \mu \nabla^2 \bv + \bff
\label{AMeq:linear}
\ ,
\eeq
where we have added the Gaussian noise term $\bff$. Since we are interested in an incompressible system, we would like the noise to be incompressible as well. In Fourier space, ${\bf f}(\bq,t ) =\int \dd^d \bbr \ee^{-\ii \bbr \cdot \bq}{\bf f}(\bbr,t )$, the incompressibility conditions implies  $\bq \cdot \bff=0$. Now, given any Gaussian noise term $\tilde{\bff}$ with statistics 
\beq
\label{eq:noiseterm}
\la \tilde{\bff} (\bbr,t) \ra = 0 \sep
\la \tilde{f}_i(\bbr,t)\tilde{f}_j(\bbr',t')\ra = 2D\delta (\bbr-\bbr') \delta (t-t') \ ,
\eeq 
we can use the transverse projection operator $P_{ij}(\bq) \equiv \delta_{ij}-q_iq_j/|\bq|^2$ to define an incompressible noise term as $f_i = P_{ij} \tilde{f_j}$.  Since $\bq \cdot \bff = q_i P_{ij} \tilde{f}_j =0$, $\bff$ is incompressible as desired. In the  Fourier transformed  space, $\bff$ has the statistics
\begin{subequations}
	\label{AMeq:noise}
	\beqn
	\la \bff(\bq,t) \ra &=& 0 
	\\
	\nonumber
	\la f_i(\bq,t)f_j(\bq',t')\ra &=& P_{ik}(\bq)P_{jh}(\bq') \la
	\tilde{f}_k(\bq,t)\tilde{f}_h(\bq',t')\ra
	\\
	&
	=&2DP_{ij}(\bq)\delta (\bq+\bq') \delta (t-t') 
	\ .
	\eeqn 
\end{subequations}
Note that the form of the noise term $\bff$ also respects all of the symmetries (symmetries (i)--(iv)) imposed on our system. Furthermore, we no longer need the Lagrange multiplier in the linear EOM  (\ref{AMeq:linear}) as it is intrinsically incompressible.

To investigate the scale invariant properties of our linear model, we now perform the following re-scaling
\beq
\bbr \mapsto \ee^{\ell} \bbr \sep \bv \mapsto \ee^{\chi \ell} \bv \sep t \mapsto \ee^{z\ell} t \ ,
\eeq
for some dimensionless number $\ell$ that describes how the spatial length scale is modified. The field variable $\bv$ and time $t$ are also re-scaled, albeit with distinct exponents: the {\it roughness exponent} $\chi$ and the {\it dynamic exponent} $z$, respectively. The numerical values of these two exponents are yet to be determined.

Applying the re-scaling to \eq (\ref{AMeq:linear}), we find
\beq
\ee^{(\chi-z)\ell} \pp_t \bv = \ee^{(\chi -2)\ell} \mu \nabla^2 \bv + \ee^{-(z+d)\ell/2}\bff
\label{AMeq:linear2}
\ ,
\eeq
where $d$ is the spatial dimension.
The prefactor in front of the noise term originates from the form of the noise term (\ref{AMeq:noise}) and the fact that the delta function scales inversely to its argument, e.g., $\delta(t) \mapsto \ee^{-z\ell} \delta (t)$. 

Re-writing \eq (\ref{AMeq:linear2}) as 
\beq
\label{AMeq:linear3}
\pp_t \bv = \ee^{(z -2)\ell} \mu \nabla^2 \bv + \ee^{(z-2\chi-d)\ell/2}\bff
\ ,
\eeq
we see that the transformed equation is exactly of the form of the original EOM (\ref{AMeq:linear}) except that the coefficients $\mu_\ell \equiv \ee^{(z -2)\ell} \mu  $ and $D_\ell =\ee^{(z-2\chi-d)\ell}D$ have acquired a dependency on $\ell$.
What it means is that if we re-scale the spatial coordinate, then the coefficients in the EOM will generically be modified. We can express the coefficients' dependencies on $\ell$ in the form of differential equations:
\beq
\frac{1}{\mu_\ell}\frac{\dd \mu_\ell}{\dd \ell} = z-2
\sep
\frac{1}{D_\ell}\frac{\dd D_\ell}{\dd \ell} = z-2\chi -d
\ .
\label{AMeq:lintheory}
\eeq
We shall call the above the {\it flow equations} of the coefficients. 

If we now pick $z$ to be 2 and $\chi$ to be $(2-d)/2$, then $\mu_\ell$ and $D_\ell$ remain unchanged as $\ell$ changes. In other words, given this choice of the exponents, the coefficients in the linear EOM are invariant under re-scaling. The beauty of this invariance is that it enables us to obtain the power-law behaviour of the temporal and spatial correlation functions of the system \cite{barabasi_b95}. For instance, we can relate the equal-time correlation function at different distance because
\beq
\la \bv(0,t) \cdot \bv(r,t) \ra = \la \bv(0,t) \cdot \bv(\ee^\ell,t ) \ra 
= \ee^{2\chi \ell} \la  \bv(0,t) \cdot \bv( 1,t) \ra  \sim r^{2\chi}
\ ,
\label{AMeq:linearscaling}
\eeq
where we have picked $\ell$ such that $\ell =\ln r$, and the second equality follows from the fact that   the re-scaling of $\bbr$ can be absorbed by re-scaling the field variable $\bv$ according to $\bv \mapsto \ee^{\chi \ell} \bv$. 

What we have seen is that in the linear theory, by suitably re-scaling the field variable and time, the coefficients in the EOM will remain invariant under spatial re-scaling, which leads to a power-law behaviour of the correlation function. Importantly, the power law exponents follow purely from the structure of the equation, and are independent of the actual coefficients in the EOM. We will now look at how the incorporation of other terms in the EOM  affects this conclusion.

\subsubsection{Nonlinear effects.}
The full EOM (\ref{AMeq:fulleom}) is a stochastic, nonlinear partial differential equation, as such, it is notoriously difficult to analyse. Here, we will employ a DRG method to treat this problem  analytically \cite{Ma1975,forster_pra77}. The strategy is to use the results from our linear theory as our reference point and then to incorporate the nonlinear effects perturbatively. To proceed, we will first employ the scaling exponents from our linear theory to gauge the importance of the additional terms in our full EOM. For example, by absorbing the scaling transformation on the term $v^4 \bv$ into the coefficient 
$c$, we have $c_\ell = \ee^{(4\chi_{\rm lin}+z_{\rm lin})\ell} c=\ee^{(6-3d)\ell} c$. If we now take $\ell$ to be big, then this term becomes small as long as the spatial dimension is greater 2. In fact, when nonlinearities are taken into account (\ref{eq:newz_chi}), this term can be shown to be negligible in the hydrodynamic limit even at $d=2$. 
Namely, if we focus on the large distance properties of the system, this particular term will  become negligible asymptotically as $\ell \rightarrow \infty$. This is also true for the $(\nabla^2)^2 \bv$ term since $\mu'_\ell = \ee^{(-4+z_{\rm lin})\ell} \mu' = \ee^{-2\ell} \mu'$. In fact, all the blue terms in   the EOM (\ref{AMeq:fulleom}) can be shown to be asymptotically negligible if $2<d<4$  according to the scaling exponents from the linear theory. We call these terms {\it irrelevant}.

How about the advective term $(\bv \cdot \nabla) \bv$? One can readily see that $\lambda_\ell = \ee^{(\chi_{\rm lin}-1+z_{\rm lin})\ell} \lambda =  \ee^{(4-d)\ell/2}\lambda$, which means that this term becomes ever more important if $d<4$ as $\ell \rightarrow \infty$. The same applies to the term $v^2 \bv$ as $b_\ell = \ee^{(2\chi_{\rm lin}+z_{\rm lin})\ell}b =\ee^{(4-d)\ell}b$.  These two terms are therefore {\it relevant} for $d<4$.

What we have seen so far is  that if $d$ is below 4, $d=3$, say, then in the hydrodynamic limit ($\ell \rightarrow \infty$), the full EOM can be reduced to 
\beq
\pp_t \bv  +\lambda (\bv \cdot \nabla ) \bv = -\nabla p + \mu \nabla^2 \bv +(a-bv^2)\bv +\bff
\ .
\label{AMeq:reducedEOM}
\eeq
Note that in contrast to the linear theory, in which the scale-invariant properties hold for any $\ell$, positive or negative, the full EOM only gets simpler as $\ell \rightarrow \infty$. In \eq (\ref{AMeq:reducedEOM}), we have also re-instated the linear term $a \bv$ as it is needed in order to fine-tune the system to criticality. The role of $a$ is similar to the role of temperature in the Ising model, which need to be fine tuned to lead the system to the critical point \cite{Cardy1996,Goldenfeld1992,Kardar2007}.

As mentioned before, using the exponents from the linear theory, the coefficients in  \eq (\ref{AMeq:reducedEOM}) only remain invariant upon re-scaling at $d=4$. For $d<4$, both the $\lambda$ and $b$ terms diverge upon zooming out spatially ($\ell \rightarrow \infty$). Looking back at \eqs (\ref{AMeq:lintheory}), we have implicitly assumed that the two flow equations are uncoupled from each other, and to find the fixed point for these two equations, it was enough to use the two free variables, $\chi$ and $z$, to make the R.H.S.~of the two equations zero. However, now that we have four coefficients in our reduced EOM ($\lambda_\ell, \mu_\ell, b_\ell$ and $D_\ell$), two variables will generically not be enough to set the  four flow equations  to zero. The problem here lies in the fact that how the coefficients vary upon increasing $\ell$ are {\it not}  independent. For instance, how $\mu_\ell$ flows can depend on the flows of $D_\ell$ and $\lambda_\ell$, etc. Realising this possibility, we  now have to analyse what kind of couplings can go into the flow equations. If we were to preserve the form of the flow equations in \eqs (\ref{AMeq:lintheory}), the R.H.S.~must be dimensionless. So the first question is to find all the dimensionless quantities that are possible out of the combinations of the coefficients. 
We now denote the dimension of spatial length $\bbr$ by $[L]$, that of time $t$ by $[T]$, and that of $\bv$  by $[U]$. Note that instead of denoting the dimension of the velocity field $\bv$ by $[L]/[T]$, we have introduced a new symbol $[U]$ to keep it general since we could have written down the EOM for the momentum field and the symmetry requirements will lead to the same identical form of EOM.
Another way to phrase it is that the mathematics encoded by the EOM does not know what the physical meaning of $\bv$ is and thus $\bv$ can be of any dimension.

From the EOM, we have
\beq
[\lambda] = \frac{[L]}{[U][T]} \ ,\ \
[\mu] = \frac{[L]^2}{[T]} \ ,\ \
[b] = \frac{1}{[T][U]^2} \ ,\ \ 
[D] = \frac{[U]^2[L]^d}{[T]} 
\ .
\eeq
Note that we have left out the coefficient $a_\ell$ associated to the term $\bv$ in the EOM. This is because $a$ is our fine-tuning parameter to take the system to criticality, and as such, the  flow of $a_\ell$ does not couple with the flows of other coefficients at the critical point \cite{chen_njp15}.

At $d=4$, there are only two dimensionless quantities that can be constructed out of these four coefficients. Two particular choices are: 
\beq
\frac{\lambda_\ell^2 D_\ell}{\mu_\ell^{3}} 
\sep
\frac{b_\ell D_\ell}{\mu_\ell^{2}} 
\ .
\label{AMeq:g0def}
\eeq
If coupling between the flow equations were to occur,  then the coupling terms can only be functions of the above two {\it coupling coefficients}. 

Going below four dimensions, the above two quantities are no longer dimensionless (because the dimension of $D_\ell$ depends on $d$), so we have to find other dimensionless coupling coefficients. In addition, we will have to write down exactly how these coupling coefficients enter the flow equations of the coefficients.
These two tasks are dealt with using a DRG transformation together with the $\epsilon$-expansion method \cite{Ma1975,forster_pra77}. Under a DRG transform, fluctuations associated with the short distance behaviour of the system are averaged over and the effects of the averaging are then incorporated back into the EOM. Since we have started with the most general EOM possible allowed by symmetry, the form of the resulting EOM must remain the same. Hence, the effects of a DRG transformation can only lead to the modifications of the coefficients in the EOM, which will be encoded in the flow equations of the coefficients. The $\epsilon$-expansion method refers to the fact that we are calculating the nonlinear effects of the system perturbatively, where the perturbation is from the linear theory (which is applicable when $d=4$), and the small parameter $\epsilon$ is the deviation from the dimension below which the linear theory breaks down, i.e., $\epsilon=4-d$ in our case. A detailed discussion of the DRG calculations for our system is beyond the scope of this review and we will only quote the results here \cite{chen_njp15}:
\begin{subequations}
	\beqn
	\frac{1}{\mu_\ell}\frac{\dd \mu_\ell}{\dd \ell} &=& z-2+\frac{1}{4}g^{(\lambda)}_\ell
	\\
	\frac{1}{D_\ell}\frac{\dd D_\ell}{\dd \ell} &=& z-2\chi -d
	\\
	\frac{1}{\lambda_\ell}\frac{\dd \lambda_\ell}{\dd \ell} &=& \chi -1+z -\frac{5}{3}g^{(b)}_\ell
	\\
	\frac{1}{b_\ell}\frac{\dd b_\ell}{\dd \ell} &=& 2 \chi +z-\frac{17}{2}g^{(b)}_\ell
	\ ,
	\eeqn
	\label{AMeq:rg}
\end{subequations}
where
\beq
\label{AMeq:defg}
g^{(\lambda)}_\ell=\frac{S_d}{(2\pi)^d}\frac{\lambda_\ell^2 D_\ell}{\mu_\ell^{3}} \Lambda^{-\epsilon}
\sep
g^{(b)}_\ell=\frac{S_d}{(2\pi)^d}\frac{b_\ell D_\ell}{\mu_\ell^{2}} \Lambda^{-\epsilon}
\ ,
\eeq
and $S_d = 2\pi^{d/2}/ \Gamma(d/2)$ is the surface area of a unit sphere in $d$ dimensions, and $\Lambda$ is some fixed short wavelength cutoff of dimension $[L]^{-1}$, whose inverse corresponds roughly to the  size of the motile organism under consideration, e.g., it can correspond to the cell diameter in a tissue. Note that the form of the coupling coefficients $g$ 
in \eq (\ref{AMeq:defg}), {\cblack obtained here from the DRG calculation,}
are the same as those in \eq (\ref{AMeq:g0def}), except for the introduction of  $\Lambda$ to keep the  $g$'s  dimensionless.

\begin{figure}
	\centering
	\includegraphics[scale=.4]{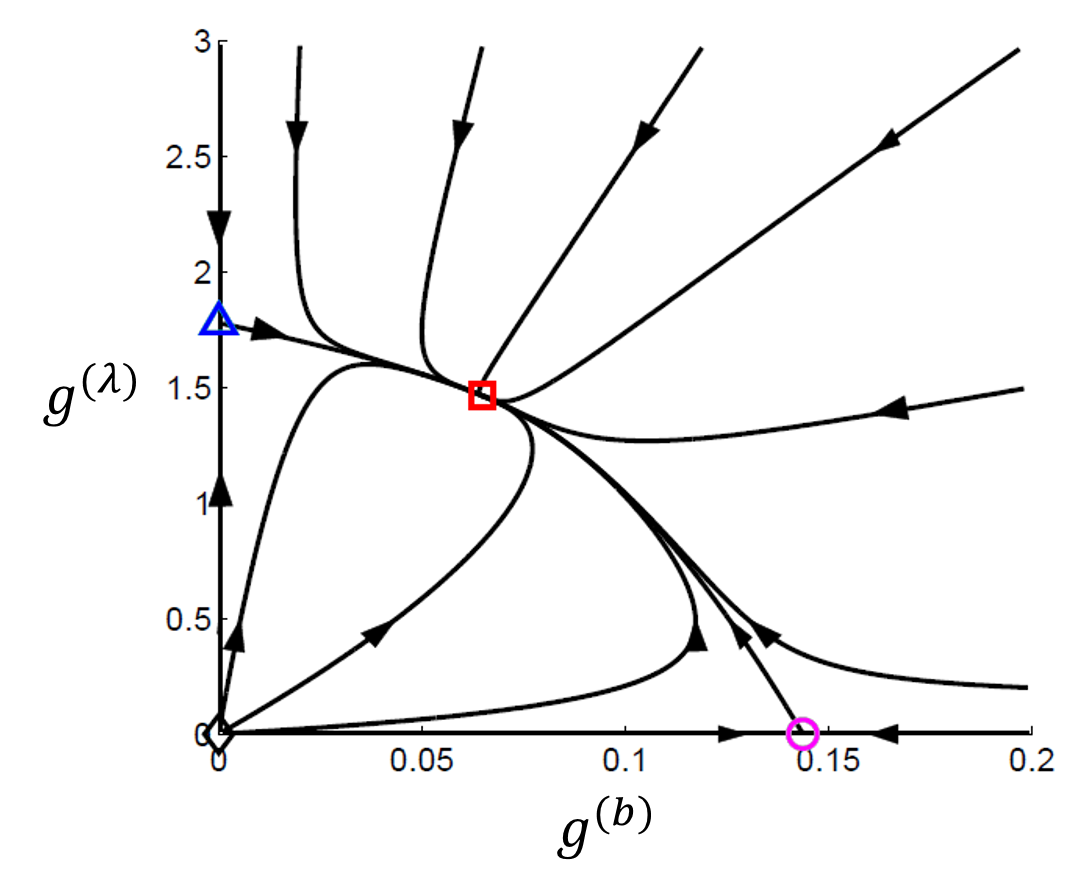}
	\caption{The flow of the two coupling coefficients $g^{(b)}_\ell$ and $g^{(\lambda)}_\ell$ under a DRG transformation for $d=3$ to order $\epsilon$. There are one stable and two unstable fixed points. The stable fixed point (red square) corresponds to the incompressible active fluid universality class found recently \cite{chen_njp15}. The unstable fixed point
		lying on the $g^{(b)}$-axis (purple circle) corresponds to the ferromagnets with dipolar interactions universality class found  in 1973 \cite{fisher_prl73}; while the		
		unstable fixed point lying on the  $g^{(\lambda)}$-axis  corresponds to the randomly stirred fluid universality class found in 1977 (model B in \cite{forster_pra77}). 	Figure reprinted from [Chen L., J. Toner and Lee C.F., New Journal of Physics, vol. 17, 042002, 2018], licensed under CC BY 3.0.
		\label{fig:RG}
	}  
\end{figure}

Since these two dimensionless quantities themselves vary with $\ell$, we can study their own flow equations. By their definitions (\ref{AMeq:defg}), we have
\begin{subequations} 	
	\label{AMeq:gflow1}
	\beqn
	\frac{1}{g^{(\lambda)}_\ell}\frac{\dd g^{(\lambda)}_\ell}{\dd \ell} &=& \frac{2}{\lambda_\ell}\frac{\dd \lambda_\ell}{\dd \ell}+\frac{1}{D_\ell}\frac{\dd D_\ell}{\dd \ell} -\frac{3}{\mu_\ell}\frac{\dd \mu_\ell}{\dd \ell}
	\\
	\frac{1}{g^{(b)}_\ell}\frac{\dd g^{(b)}_\ell}{\dd \ell} &=& \frac{1}{b_\ell}\frac{\dd b_\ell}{\dd \ell}+\frac{1}{D_\ell}\frac{\dd D_\ell}{\dd \ell} -\frac{2}{\mu_\ell}\frac{\dd \mu_\ell}{\dd \ell}
	\ .
	\eeqn
\end{subequations}
Using the expressions in the flow equations (\ref{AMeq:rg}), we can re-write the R.H.S.~above as
\begin{subequations} 	
	\label{AMeq:gflow2}
	\beqn
	\frac{1}{g^{(\lambda)}_\ell}\frac{\dd g^{(\lambda)}_\ell}{\dd \ell} &=& \epsilon-\frac{3}{4}g^{(\lambda)}_\ell -\frac{10}{3} g^{(b)}_\ell
	\\
	\frac{1}{g^{(b)}_\ell}\frac{\dd g^{(b)}_\ell}{\dd \ell} &=& \epsilon-\frac{17}{2}g^{(b)}_\ell -\frac{1}{2} g^{(\lambda)}_\ell
	\ .
	\eeqn
\end{subequations}
\eqs (\ref{AMeq:gflow2}) now define two coupled ordinary differential equations, and their dependencies on $\ell$ are shown in \fig \ref{fig:RG}. As $\ell \rightarrow \infty$, the coupling coefficients  flow to three fixed points. Two are unstable (depicted by the purple circle and the blue triangle in \fig \ref{fig:RG}) and were already known to physicists in the context of ferromagnetism with dipolar interactions  \cite{fisher_prl73, aharony_prb73,aharony_prb73b} and randomly stirred fluids (model B in \cite{forster_pra77}). The stable fixed point (red square), in the sense that all flow lines around the fixed point converge to it as $\ell$ increases, was a novel addition whose discovery  was solely motivated by the study of active matter \cite{chen_njp15}. We will now focus on this fixed point.

At the stable fixed point, 
\beq
g^{(\lambda)}_{\infty} = \frac{123}{113} \epsilon +{\cal O}(\epsilon^2)
\sep
g^{(b)}_{\infty} = \frac{6}{113} \epsilon +{\cal O}(\epsilon^2)
\ ,
\eeq
and the R.H.S. of \eqs  (\ref{AMeq:gflow2}) are zero and thus enforce two algebraic  constraints on the flow equations of the coefficients (\ref{AMeq:rg}) via \eqs (\ref{AMeq:gflow1}). If we now pick $z$ and $\chi$ such that two of the flow equations are zero, those of $\mu_\ell$ and $D_\ell$, say, then together with the two previous constraints, they imply the invariance of all coefficients as $\ell \rightarrow \infty$. In this asymptotic regime, the system will again exhibit scale invariant properties as in the linear theory, except now, due to the nonlinearities in the system, the roughness and dynamics exponent are modified to
\beq
\label{eq:newz_chi}
z =2-\frac{31}{113} \epsilon \sep \chi = \frac{2-d}{2} -\frac{31}{226} \epsilon
\ .
\eeq
In particular, in contrast to the scaling predicted by the linear theory in \eq (\ref{AMeq:linearscaling}), when the nonlinear effects are taken into account, the equal-time correlation of the system actually follows  the power-law:
\beq
\la \bv(0) \cdot \bv(r) \ra \sim r^{2\chi} = r^{2-d-31\epsilon/113}
\ .
\eeq
Note that we focus here on the system's behaviour right at the critical point so that the correlation length of the system  is infinite. Away from criticality, there is  another exponent that quantifies how the correlation length varies as one gets close to the critical point. We refer the interested readers to \cite{chen_njp15} for further details.

\subsubsection{Universality class.}
We have seen that by a symmetry consideration alone, we can derive a model EOM that generically describes incompressible active fluids. By focusing on the large-distance behaviour (the limit $\ell \rightarrow \infty$), we have managed to calculate the scaling behaviour of the system in spatial dimensions $4-\epsilon$. Importantly, at no point did we need the inputs of the actual model parameters. Therefore, the exponents governing the scaling behaviour depend only on the symmetry of the system, and are otherwise oblivious to the quantitative parameters that describe the actual systems. This suggests that a large class of dynamical systems having the same form of EOM, but with distinct model parameters, will exhibit identical scale invariant behaviour. These distinct systems are  said to belong to the same {\it universality class}.

\subsection{Ordered phase in two dimensions}
\label{AMsec:ordered}
We have seen that at the critical transition, the scaling behaviour of a generic incompressible active fluid constitutes a novel universality class in non-equilibrium physics. Here, we will describe how in two dimensions, the ordered phase in incompressible active fluids also exhibits universal  behaviour, albeit with scaling behaviour that belongs to a well known universality class: the Kardar-Parisi-Zhang ({\bf KPZ}) universality class that originated from modelling surface growth in the non-equilibrium regime \cite{kardar_prl86}.

\subsubsection{Linear theory.}
As before, the symmetry consideration alone has fixed the structure of the EOM. The difference  from the previous section is that while at criticality, $v$ is small and thus an expansion with respect to $\bv$ makes sense, this is no longer true in the ordered phase since $v$ can be large. As a result, the generic EOM in the ordered phase is different from the EOM in \eq (\ref{AMeq:reducedEOM}) \cite{chen_natcomm16}. We will however continue to use the restricted form (\ref{AMeq:reducedEOM})  for simplicity and our conclusion remains the same even for the generic case \cite{chen_natcomm16}. 

Without loss of generality, we can assume here that the collective motion is along the $x$-direction and re-write the field variables $\bv$ in terms of the deviation field $\bu$ from the mean velocity $|\la\bv\ra|=v_0$:
\beq
\bv = (v_0+u_x)\hat{\bf x} +u_y  \hat{\bf y}
\ ,
\eeq
and we assume $u \equiv |\bu|$ is much smaller than $v_0$.

Re-writing the EOM in terms of $\bu$ and then keeping only the leading order linear terms, we have
\beq 
\pp_t \bu  +\lambda v_0 \pp_x \bu = -\nabla p + \left( \mu_x\nabla^2 u_x -2a u_x\right)  \hat{\bf x} +\mu_y \nabla^2 u_y \hat{\bf y}+\bff
\label{AM:eq_lineq}
\ ,
\eeq
where $v_0= \sqrt{a/b}$, and 
$p(\bbr,t)$ again serves to impose the incompressibility condition $\nabla \cdot \bu =0$. Specifically, we have
\beq
\nabla^2 p = \pp_x \left( \mu_x\nabla^2 u_x -2au_x\right)   + \pp_y ( \mu_y \nabla^2  u_y)
\ .
\eeq
In the Fourier-transformed space,
\beq
-q^2 p = -\ii q_x \left( -\mu_x q^2 u_x -2a u_x\right)  + \ii q_y q^2 \mu_y   u_y
\label{eq:expp}
\ ,
\eeq
where $q \equiv | \bq|$.

In \eq (\ref{AM:eq_lineq}), we have also allowed for distinct ``viscosity'' coefficients $\mu_{x,y}$ for the two directions since in the ordered phase, the rotational symmetry is broken and there is no reason to expect that the scaling behaviour will be identical in both $x$ and $y$ directions. For the same reason, we will allow for two distinct direction-dependent roughness exponents $\chi_x$ and $\chi_y$, as well as an {\it anisotropic} exponent $\zeta$ in our re-scaling scheme:
\beq
\label{AMeq:orderedscaling}
x \mapsto \ee^{\ell} x \sep y \mapsto \ee^{\zeta \ell} y \sep u_x \mapsto \ee^{\chi_x \ell} u_x \sep u_y \mapsto \ee^{\chi_y \ell} u_y \sep t \mapsto \ee^{z\ell} t \ .
\eeq
Although there are now seemingly five distinct exponents, they are not all independent as some of the exponents are related via the incompressibility condition: $q_x u_x= -q_y u_y \mapsto \ee^{(\chi_x-1)\ell} q_x u_x= -\ee^{(\chi_y-\zeta)\ell} q_y u_y$, which implies that in the hydrodynamic limit, 
\beq
\label{AMeq:chirelation}
\chi_x=\chi_y-\zeta+1
\ .
\eeq

In two dimensions, the ``potential energy'' term $\cH$ is depicted in \fig \ref{fig:AMphase}, whose functional derivative with respect to $\bv$ gives rise to the terms $(a+b^2v^2)\bv$ in the EOM. When the symmetry is broken, we know from equilibrium physics that fluctuations in $v_y$ dominate over those in  $v_x$ because moving along the trough of $\cH$ costs much less energy than moving up the valley in $\cH$. We assume the same principle applies to our non-equilibrium problem here, which we will verify a posteriori. 

Focusing therefore on $u_y$, we first go to the boost frame $\bu(x,y,t) \mapsto \bu(x-\lambda v_0t,y,t)$ to eliminate the second term on the L.H.S.~of \eq (\ref{AM:eq_lineq}). Without this term, we have
\begin{subequations}
	\beqn
	\pp_t u_y  &=& \ii q_y p -\mu_y q^2 u_y +f_y
	\\
	\label{AM:eqorderedlin}
	&=&  \left(\mu_x +\frac{2a}{q^2}\right)q_xq_yu_x -\mu_y q_x^2 u_y +f_y
	\ ,
	\eeqn
	where (\ref{AM:eqorderedlin}) follows from (\ref{eq:expp}).
	
\end{subequations}
We now use the incompressibility condition $q_xu_x+q_yu_y=0$ to express $u_x$ as $-q_yu_y/q_x$. \eq (\ref{AM:eqorderedlin}) can thus be written as 
\beq
\pp_t u_y  = -\left[  \mu_xq_y^2 +2a\frac{q_y^2}{q^2} +\mu_y q_x^2\right] u_y +f_y
\ .
\eeq
Upon performing the re-scaling as in \eq (\ref{AMeq:linear3}), we have
\beq
\pp_t u_y=- \ee^{(z-2\zeta)\ell} \mu_xq_y^2u_y -\ee^{(z-2\zeta+2)\ell}2a\frac{q_y^2}{q^2}u_y -\ee^{(z-2)\ell}\mu_y q_x^2u_y  
+
\ee^{(z-2\chi_y-1-\zeta)\ell/2}
f_y
\label{AM:eqlinscaling}
\ .
\eeq
We have argued previously that fluctuations in $u_y$ dominate over those of $u_x$, we thus expect that $\chi_y> \chi_x$, this in turn leads to the inequality $\zeta>1$ via \eq (\ref{AMeq:chirelation}). Therefore, $q^2 \mapsto \ee^{-2\ell} q_x^2 +\ee^{-2 \zeta \ell} q_y^2 \sim \ee^{-2\ell}q_x^2$. This explains the prefactor in the second term on the R.H.S. of \eq (\ref{AM:eqlinscaling}). Furthermore,  we can conclude that $\mu_x q_y^2 u_y$, the first term in the R.H.S.~of \eq (\ref{AM:eqlinscaling}), is not as important as the third term upon coarse-graining. Ignoring the first term, we can make the remaining three terms invariant upon re-scaling by choosing the following exponents:
\beq
\label{AMeq:linorder}
\zeta_{\rm lin}= 2 \sep \chi_{y, {\rm lin}} =- \frac{1}{2} \sep z_{\rm lin}= 2
\ .
\eeq
In addition, \eq (\ref{AMeq:chirelation}) implies $\chi_{x, {\rm lin}}=-3/2$.

\subsubsection{Nonlinear effects.}
Going back to the simplifed EOM (\ref{AMeq:reducedEOM}), we can now gauge the relevance of the two distinct nonlinear terms. We  consider first the advective term: $\lambda (\bv \cdot \nabla) \bv$. Focusing again on the EOM of $u_y$, the nonlinearities coming from this term are
$\lambda u_x \pp_x u_y$ and $\lambda u_y \pp_y u_y$. Upon re-scaling, these terms become
$\ee^{(z+\chi_x-1)\ell}\lambda u_x \pp_x u_y$ and $\ee^{(z+\chi_y-\zeta)\ell}\lambda u_y \pp_y u_y$. Using the exponents from the linear theory, we can see that both terms are irrelevant as $\ell \rightarrow \infty$. On the other hand, the nonlinearities associated to the coefficient $a$ are of the form
\beq
-\frac{2a}{v_0} u_xu_y -\frac{a}{v_0^2}u_x^2u_y -\frac{a}{v_0^2} u_y^3 
\ ,
\eeq
which, upon re-scaling, becomes
\beq
- \ee^{(z+\chi_x)\ell} \frac{2a}{v_0} u_xu_y - \ee^{(z+2\chi_x)\ell}\frac{a}{v_0^2}u_x^2u_y - \ee^{(z+2\chi_y)\ell}\frac{a}{v_0^2} u_y^3
\ .
\eeq
In this case, both the first and third terms are relevant and we will thus keep these two terms. 

In summary, for an incompressible active fluid in two dimensions, the governing EOM (\ref{AMeq:reducedEOM}) can be further simplified, in the hydrodynamic limit, to 
\beq
\label{AMeq:reducedordered}
\pp_t \bu   = -\nabla p +\mu \pp_x^2 u_y \hat{\bf y} -a\left(\frac{2u_x}{v_0}+\frac{u_y^2}{v_0^2}\right)(v_0\hat{\bf x}+u_y\hat{\bf y})  +\bff
\ ,
\eeq
where we have omitted the subscript $y$ in $\mu$ above to ease notation. Note that we have not considered the role of the pressure term $p$ in detail in the above analysis, however our conclusion remains 
the same even if we do so \cite{chen_natcomm16}.

\subsubsection{Mapping active fluids in two dimensions onto an equilibrium system.}
{\cblack  To make analytical progress with the EOM (\ref{AMeq:reducedordered}), we will first map the class of systems described by (\ref{AMeq:reducedordered}) onto an equilibrium system. }
As the advective term $\lambda (\bv \cdot \nabla) \bv$ is irrelevant in the ordered phase, its omission enables us to re-write the hydrodynamic EOM \eq (\ref{AMeq:reducedordered}) as
\beq
\label{AMeq:reducedordered2}
\pp_t \bu  = -\frac{\delta \cH}{\delta \bu} -\nabla p+\bff 
\ ,
\eeq
where 
\beq
\cH =  \int \dd^2 \bbr \left[a \left(u_x +\frac{u_y^2}{2v_0}\right)^2 +\frac{\mu}{2}(\pp_x u_y)^2 \right]\ .
\eeq
Viewing $\cH$ as the Hamiltonian of the system, 
{\cblack and the noise term $\bff$  as thermal fluctuations,} we can analyse the static, or equal-time, properties of the system  by studying the corresponding partition function:
\beq
Z = \int {\cal D}^2 \bu \delta(\nabla \cdot \bu) \ee^{-\beta \cH[\bu]}
\ ,
\eeq
where $\beta = 1/D$ {\cblack (\ref{eq:noiseterm})} and ${\cal D}^2 \bu \equiv  \lim_{N\rightarrow \infty} \prod_{k}^N \dd u_x(\bbr_k,t_k) \dd u_y(\bbr_k,t_k)$, with the index $k$ enumerating the discretised but infinitesimal elements of both space and time. Note that the incompressibility constraint $\nabla \cdot \bu=0$ is enforced by the delta function $\delta(.)$ in the partition function above.

\begin{figure}
	\centering
	\includegraphics[scale=.5]{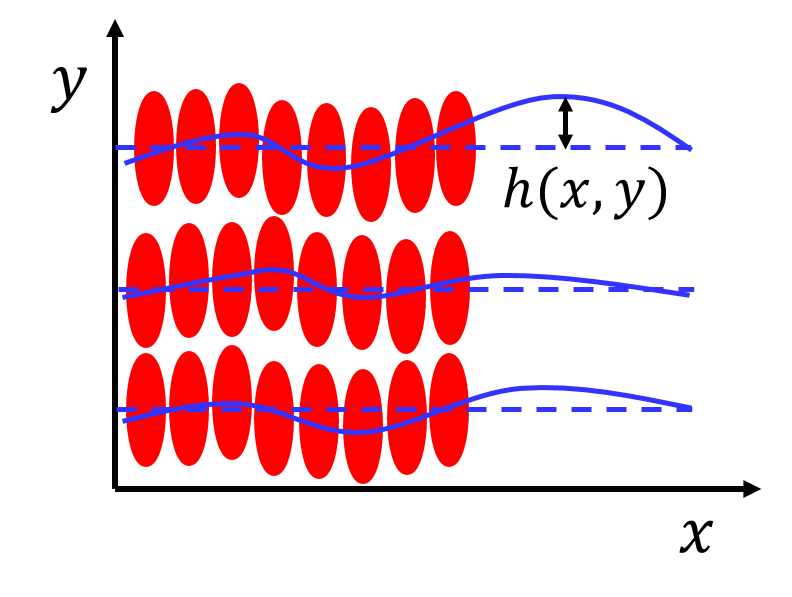}
	\caption{
		A smectic liquid crystal is a  layered structure of liquid crystals (red ellipsoid) and the function $h(x,y)$ quantifies the deviation of the layer at position $(x,y)$ from the expected location.
		\label{fig:AMsoap}
	}  
\end{figure}

Since we are in two dimensions, we can further eliminate the incompressibility constraint by using the stream function $h(x,y)$ defined as
\beq
\label{AMeq:stream}
u_x =-v_0\pp_y h \sep u_y= v_0\pp_x h
\ .
\eeq 
In terms of $h$, the Hamiltonian $\cH$ becomes
\beq
\cH_{\rm S} =  \int \dd^2 \bbr \left[\frac{B}{2} \left(\pp_y h -\frac{(\pp_x h)^2}{2}\right)^2 +\frac{K}{2} \left(\pp_x^2h \right)^2 \right]
\ ,
\eeq
which is exactly the Hamiltonian that describes a dislocation-free {\it smectic liquid crystal} in two dimensions (\fig \ref{fig:AMsoap}),
with $B \equiv 2av_0^2$ and $K\equiv\mu v_0^2$ being the compression modulus and the bending modulus, respectively  \cite{chaikin_b00}.  
In a smectic liquid crystal, the liquid crystals (depicted as red ellipsoids in \fig \ref{fig:AMsoap}) formed a layered structure in which the layers are parallel to $x$-axis on average and $h(x,y)$ describes the height deviation of the layers from the expected location. The word smectic comes from the Greek word for soap, whose slippery surface upon lubrication consists of layered lipid bilayers which can slide across each other easily.

Note that since $\cH_{\rm S}$ has to be invariant under a rotation with respect to the axis perpendicular to the $xy$-plane,  the compression term should also include the term $-(\pp_y h)^2/2$ in the curly brackets. However, such a term is irrelevant in the hydrodynamic limit and thus omitted.

Given the  Hamiltonian $\cH_{\rm S}$, at thermal equilibrium, the probability of having a particular profile $h(x,y)$ is given by
\beq
{\rm Prob.}[h] =\frac{ \ee^{-\beta \cH_{\rm S} [h]}}{Z_{\rm S}} \ ,
\label{AMeq:h}
\eeq
where the partition function is 
\beq
\label{AMeq:zsdef}
Z_{\rm S} = \int {\cal D} h \ee^{-\beta \cH_{\rm S}[h]}
\ .
\eeq

\subsubsection{Kardar-Parisi-Zhang universality class.} 
Besides the surprising connection between incompressible active fluids and smectics in two dimensions, we will now make one further connection to a well known physical model: the KPZ surface growth model, based on a mapping devised in \cite{Golubovic1992,Golubovic1994}. 
To do so, we first add the two boundary terms below (hence immaterial due to the assumed periodic boundary condition) to the smectics Hamiltonian:
\begin{subequations}
	\beqn
	0=\tri \cH_{\rm S} &=& \frac{BK}{2}\int \dd^2 \bbr \left\{\pp_y  \left[(\pp_x h)^2\right] +
	\pp_x \left[\frac{1}{6}(\pp_x h)^3\right]
	\right\}
	\\
	&=&\frac{BK}{2}  \int \dd^2 \bbr \left[  (\pp_x h) (\pp_{xy} h) +
	\frac{1}{2} (\pp_x h)^2 (\pp_x^2 h)
	\right]
	\\
	&=&  \frac{BK}{2}  \int \dd^2 \bbr \left[ - (\pp_y h) (\pp_{x}^2 h) +
	\frac{1}{2} (\pp_x h)^2 (\pp_x^2 h)
	\right]
	\ ,
	\eeqn
\end{subequations}
where the last equality follows from an integration by parts on the first term.

Adding these immaterial boundary terms to the smectics Hamiltonian enables us to ``complete the square'' in the integrand and to re-write $\cH_{\rm S}$ as
\beq
\cH_{\rm S} =  \int \dd^2 \bbr \left[\frac{B}{2}\left(\pp_y h -\frac{(\pp_x h)^2}{2}\right) +\frac{K}{2} \left(\pp_x^2h \right) \right]^2
\ .
\eeq
Employing the delta function $\delta[.]$, we can now re-write the partition function (\ref{AMeq:zsdef}) as
\beq
Z_{\rm S} = \int {\cal D}h \left\{ \int {\cal D} \eta\ee^{-\beta\int \dd^2 \bbr \dd t \eta(\bbr, t)^2}
\delta \left[ \frac{B}{2}\left(\pp_y h -\frac{(\pp_x h)^2}{2}\right) +\frac{K}{2} \left(\pp_x^2h \right)-\eta\right] \right\}
\label{AMeq:Zs}
\ ,
\eeq
where $\eta$ may be viewed as an auxilliary field to be integrated over.

However, another way to interpret $\eta(\bbr, t)$ is to view it as a random variable for any particular $\bbr$ and $t$, such that its probability density is proportional to $\ee^{-\beta \eta (\bbr, t)^2}$. 
From this perspective, we can generate the same statistics of height profile $h$ as in  \eq (\ref{AMeq:h}) by using the  Langevin equation:
\beq
\frac{B}{2}\pp_y h = \frac{K}{2} \pp_x^2 h +\frac{B}{4} (\pp_x h)^2 +\eta
\label{AMeq:kpz}
\ ,
\eeq
where
\beq
\la \eta(x,y) \ra = 0 \sep
\la \eta(x,y)\eta(x',y')\ra = 2D\delta (x-x') \delta (y-y') \ .
\eeq
\eq (\ref{AMeq:kpz}) is exactly the (1+1)$d$ Kardar-Parisi-Zhang (KPZ) surface growth model if one interprets the symbol $y$ as time and $x$ as the one spatial dimension 
in the KPZ equation.

\begin{figure}
	\centering
	\includegraphics[scale=.6]{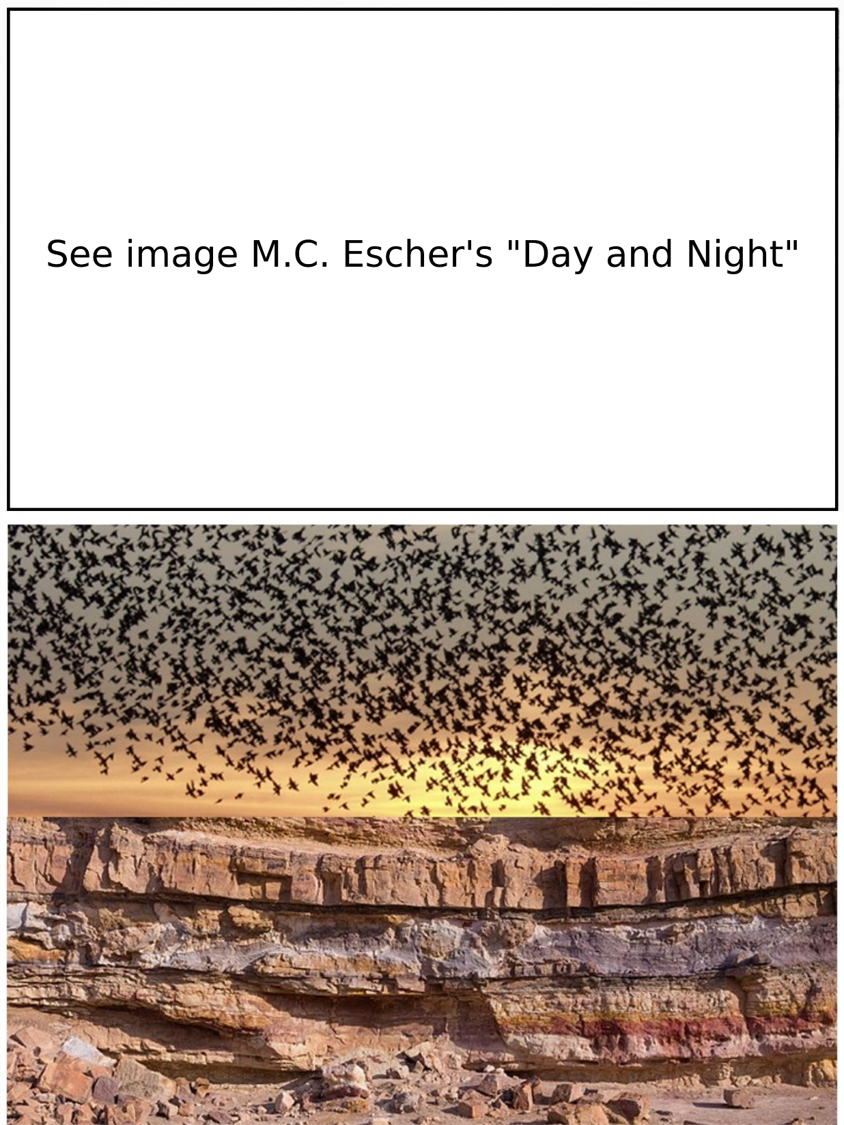}
	\caption{Our discussion in Sect.~\ref{AMsec:ordered} demonstrates that in two spatial dimensions, the large distance properties of an incompressible active fluid in the ordered phase, e.g., incompressible flock, has the same scale-invariant properties as a two-dimensional smectic liquid crystal 
		and the KPZ surface growth model. The findings thus suggest
		that instead of the apparent equivalence between an incompressible
		flock and a two-dimensional lattice depicted by the landscape in 
		M.C. Escher's ``Day and Night'',
		the flock should instead 
		be mapped onto a smectic-like 
		layered structure such as sedimentary rocks \cite{cuerno_interface12} [Top half of bottom figure: photograph by Tommy Hansen. 
		Public domain.
		Bottom half of bottom figure: photograph by Ryan McGrady, 
		distributed under a CC-BY 4.0 license.]. 
		\label{fig:escher}
	}  
\end{figure}

The roughness and dynamic exponents of the KPZ model are known exactly: $z_{\rm KPZ} = 3/2$ and $\chi_{\rm KPZ} = 1/2$ \cite{kardar_prl86}. Translating this back to our active fluid model, the KPZ dynamic exponent becomes the anisotropic exponent: $\zeta = 3/2$. Furthermore, 
via the relationships between $\bu$ and $h$ in \eq (\ref{AMeq:stream}), we have
\beq
\chi_x=\chi_{\rm KPZ} -\zeta = -1
\sep
\chi_y = \chi_{\rm KPZ} -1 = -\frac{1}{2}
\ .
\eeq 
Compared to the exponents obtained from our linear theory (\ref{AMeq:linorder}), we see that the anisotropic exponent $\zeta$, and as a result also the roughness exponent along the $x$ direction $\chi_x$, are modified due to the nonlinear terms in the EOM. 
Furthermore, since $\chi_y > \chi_x$, we can now verify the previous assertion that $u_y$ dominates over $u_x$ in the hydrodynamic limit 
(\fig \ref{fig:escher}).

Looking at the re-scaling scheme in (\ref{AMeq:orderedscaling}), the only exponent that remains unknown is the dynamic exponent $z$. As we have mapped the active fluids onto an equilibrium system and then focused on the partition function, the connection established actually does not allow us to determine the dynamic exponent $z$. Hence, what the value of $z$ is remains an interesting open question.

\section{Conclusion \& Outlook}
The behaviour of a system undergoing a phase transitions is an archetype of collective phenomena, and the emergence of universality at a phase transition makes it a fascinating subject for physicists. Motivated by recent studies focused on phase transitions in  biological systems, we have discussed how novel physics can arise from the generic non-equilibrium nature of living matter, be it driven chemical reactions or self-generated mechanical forces.  For the former, we have seen in Sect.~2 how driven chemical reactions in the cell cytoplasm can stabilise protein drops that form by phase separating out of the cytosol, contrary to the universal coarsening behaviour found in phase separating systems at thermal equilibrium. In Sect.~3, we focused on incompressible active matter and we have shown that a novel universality class emerges at the critical order-disorder transition due to the activity of the system. Furthermore, in the ordered phase in two dimensions, we have discussed the  surprising connections between active fluids, smectics, and the Kardar-Parisi-Zhang surface growth model.

Besides the two particular systems discussed in this review, there are  other examples of phase transition phenomena in biology that have led to the discovery of novel physics. Here, we will mention two particular examples that are receiving intense attention from physicists.
\begin{enumerate}
	\item{\it Motility-induced phase separations.}
	Motivated by the run-and-tumble motion of bacteria \cite{tailleur_prl08}, the study of active particles that interact solely via volume exclusion led to the discovery of liquid-gas phase separation   driven purely by motility \cite{fily_prl12, redner_prl13,
		Cates2015,cates_annrev18}. The condensed drops in motility-induced phase separations (MIPS) show interesting interfacial properties, such as the existence of a stable interface with a negative surface tension  \cite{bialke_prl15}. Away from the critical point and close to the phase boundary, the coarsening kinetics of MIPS has been argued to be identical to the Liftshitz-Slyozov scaling law found in equilibrium system  \cite{lee_softmatter17}. However, what the universal behaviour of MIPS is at criticality remains an interesting open question.
	\item{\it Active polymer networks.}
	Throughout the review, the non-equilibrium processes, be they active motion or chemical reactions, occur in a homogeneous environment that corresponds to an Euclidean space. What happens if active forces are now transmitted through a network of irregular structures instead? A recent discovery found  that a biologically relevant active polymer network under fragmentation can self-organise itself to exhibit a scale-invariant signature of a critical system \cite{Alvarado2013,alvarado_softmatt17}. While the exponents observed are close to that of the static percolation universality class \cite{Sheinman2015,lee_pre16}, the question of whether the critical phenomenon in active networks actually belongs to the static percolation universality class remains unsettled \cite{Sheinman2015a,pruessner_prl16,sheinman_prl16,hu_prl16}.
\end{enumerate}

\vspace{.2in}
\noindent
In terms of outlook, we believe the following future directions will expand the horizon of both biology and physics.
\begin{enumerate}		
	\item
	In \Sec 2 we have studied how driven chemical reactions can stabilise a multi-drop, ternary system. As the cell cytoplasm is a complex mixture of thousands of different molecules \cite{sear_prl03, jacobs_biophysj17} it will be interesting to see how these results may be modified in many-component mixtures.
	
	\item In \Sec 2.4 we have provided intuitive arguments 
	to explain the appearance of a lattice structure of phase-separated drops in our Monte Carlo simulations. Such a structure naturally suggests a kind of repulsive interactions between drops, which may serve to stabilise a multi-drop system against coarsening via coalescence due to drop diffusion. Whether this is indeed the case remains to be investigated.
	\item
	In Sect.~3, we have studied the simplest kind of symmetry: the rotational symmetry and the associated universal behaviour when the symmetry breaks spontaneously in an active system. But what are the other relevant symmetries in biology, and will they also lead to novel universal  behaviours?
\end{enumerate}

\appendix


\section{Non-equilibrium phase separation in the small drop regime: solute concentration profile in the cytoplasm}
\label{sec:app:cyto}
We analyse the cytoplasmic solute concentration $\Po(r)$ (\eq \eqref{profileFull}) in the small drop regime ($R\ll\xi$, \eq \eqref{xi}), in two limiting cases. First, we assume the system contains few drops so that the inter-drop distance is large compared to the gradient length scale $\xi$. The term $U_{\rm in}^{(1)}$ is zero to avoid a diverging concentration far from drops ($r \gg \xi$), and we fix the interface concentration ($r=R$) according to the Gibbs-Thomson relation \eq \eqref{gtP}:
\beqn
\label{Po_small}
\Po(r) &=& \Po(\infty) + \left(\Po(R) - \Po(\infty)\right)  \frac{R}{r} \ee^{-(r-R)/\xi} \ .
\eeqn
where $\Po(\infty)$ is the concentration far from drops. We now assume that many drops are present so that the inter-drop distance is small compared to $\xi$. The quantity $r / \xi$ is then always small and the exponential terms in \eq \eqref{profileFull} are close to one. Imposing again the Gibbs-Thomson relation at the interface we find $\Po(r) = \Pinf + (\Po(R)-\Pinf) R/r$, which is equivalent to \eq \eqref{Po_small} for vanishing $r/\xi$. Therefore \eq \eqref{Po_small} is a good approximation for the cytoplasmic profile $\Po(r)$ in both regimes of small and large drop number.

The cytoplasmic profile $\Po(r)$ is closely similar to that at equilibrium (i.e without chemical reactions, \Sec \ref{sec:equi}), for identical supersaturation $\Delta$. This can be seen by comparing \eqs \eqref{profileEq} and \eqref{Po_small} in the vicinity of and far from the drop interface ($r \approx R$ and $r \gg R$). The influx of solute entering the drop, $J_{\rm out \rightarrow in} \equiv D \dd \Po/ \dd r|_{r=R}$ is
\beqn
J_{\rm out \rightarrow in} &=& \frac{D}{R}\left( 1 + \frac{R}{\xi} \right) \left(\Delta - \frac{\hatPo l_c}{R} \right) \ .
\eeqn
Since $R \ll \xi$ is small by definition in the small drop regime we recover the solute flux in absence of chemical reactions \eq \eqref{Jeq}.

\section{Non-equilibrium phase separation in the large drop regime}
\label{sec:app:large}

In the large drop regime the drop radii $R$ are large compared to the gradient length scale $\xi$ (\eq \eqref{xi}). The concentration profiles (\eqs \eqref{profileLargeIn} and \eqref{profileLargeOut}) are shown in \fig \ref{fig:noneq_large1} and the molecular fluxes at the drop interfaces are given by \eqs \eqref{Joi_large} and \eqref{Jio_large}. We will first study the system at the steady state, all drops having the same radius $R^*$, and then we will analyse its stability against Ostwald ripening.

\subsection{Steady state}
\label{sec:app:large:steady}

Besides our assumption that $S$ does not phase separate i.e. $\Sin(R)=\So(R)$ (see Ref \cite{wurtz_prl18} for the more general analysis without this assumption), we enumerate the other constraints on the system at steady state:

\begin{enumerate}

	\item
	{\bf The Gibbs-Thomson relations dictate the concentrations at the interfaces} (\GT), which follow from the assumption that the system is close to equilibrium to the extent that local thermal equilibrium is true (see main text).
	
	\item
	{\bf Drops and cytoplasm are at chemical equilibrium:} in this regime drops are large compared to the gradient length scale $\xi$, and since we only focus on low drop density systems the same is true for the cytoplasm. Therefore in both phases and far away from interfaces the concentration profiles must be flat. Taking $\nabla^2 P = \nabla^2 S = 0$ in the reaction-diffusion equations \eqs \eqref{reacdiffP} and \eqref{reacdiffS} it follows that the concentrations are at chemical equilibrium. This imposes the following constraints on the solute concentration in the drop centre ($\Pin(0)$) and far from drops ($\Pinf$):
	\beqn
	\label{eq:cons3}
	\Pin(0) = \frac{h}{k} \Sin(0) \sep  \Pinf = \frac{h}{k}\So(\infty) \ .
	\eeqn

	\item
	{\bf The solute mass conservation} imposes the relation:
	\beqn
	\label{eq:cons2}
	N \frac{4 \pi R^3}{3} \Pin(0) + \left(V-N \frac{4 \pi R^3}{3} \right) \Po(\infty) = V \Ptot \ ,
	\eeqn
	where $N$ in the number of drops, $V$ the total volume of the system and $\Ptot=\phi h/(k+h)$ is the global concentration of $P$ in the entire system \cite{wurtz_prl18}.	Note that we have neglected the concentration gradients near the interface due to the relative small size of the interfacial region ($\sim \xi$) compared to the size of the drops and the cytoplasm in this regime.

	\item
	{\bf The combined concentration $P+S$ is homogeneous in drops and in the cytoplasm}, as already shown in the main text, \eq \eqref{P+S}. This allows to express the concentrations $\Pin(0),\Sin(0)$ in the drop centre and $\Pinf, S_{\rm out}(\infty)$ far from drops as functions of the interface concentrations:
	\beqn
	\Pin(0) + \Sin(0) &=& \Pin(R) + \Sin(R) \\
	\Pinf +S_{\rm out}(\infty) &=& \Po(R) + \So(R)
	\ .
	\eeqn
	
	\item
	{\bf The flux balance condition}: the fluxes at the drop interfaces $\Joi$ and $\Jio$ (\eqs \eqref{Joi_large} and \eqref{Jio_large}) must balance each other at steady state so that drops neither grow nor shrink:
	\beqn 
	\label{eq:cons1}
	\Po(\infty) - \Po(R) = \Pin(R) - \Pin(0) \ ,
	\eeqn
	where we have assumed infinite drop radius ($\xi/R \rightarrow 0$).
\end{enumerate}

Note that the conditions from (i) to (iv) also apply when drop radii are not at steady state because we assumed that the concentration profiles are always at the steady state ($\partial P/\partial t = \partial S/\partial t = 0$, \Sec \ref{sec:equi}). Imposing all constraints above we can determine the steady state drop radius $R^*$ \cite{wurtz_prl18}:
\beqn
N\frac{ 4 \pi (R^*)^3}{3} &=&  \left( \frac{\phi - \hatPo}{\hatPin} - \frac{k}{2 h} \right) V \ .
\eeqn
Contrary to the small drop regime (\Sec \ref{sec:small}) the drop radius $R^*$ scales with the system size $V$, as in equilibrium systems i.e., systems without chemical reactions. This result shows that there exists a maximal forward rate constant $k_u$ above which drops dissolve ($R^*=0$) \cite{wurtz_prl18}:
\beqn
\label{app:ku}
k_u =  2\frac{\phi - \hatPo}{\hatPin} h \ .
\eeqn
For $k>k_u$ drops cannot exist in the large drop regime but may still exist in the small drop regime (\Sec \ref{sec:small}).

\subsection{Stability of the steady state against Ostwald ripening}
\label{sec:app:large:stability}

Let us now analyse the stability of our system against Ostwald ripening. We consider a multi-drop system initially at steady state, all drops radii being $R^*$ (\ref{sec:app:large:steady}). The molecular fluxes at a drop interface are given by \eqs \eqref{Joi_large} and \eqref{Jio_large}. We now perturb the system by randomly increasing or decreasing each drop radius $R$ by a fixed quantity  $\epsilon$, much smaller than $R^*$. Let us concentrate on a specific drop $i$ whose radius $R_i$ is increased: $R_i = R^* + \epsilon$. The new interface fluxes for this drop are (\eqs \eqref{Joi_large} and \eqref{Jio_large})
\beqn
\label{app:Joi_large}
\Joi &=&   \frac{D  \left( \Pinf - \Po(R_i) \right)}{\xi}  \left( 1 + \frac{\xi}{ R_i} \right) \\
\label{app:Jio_large}
\Jio &=& \frac{D \left( \Pin(R_i) - \Pin(0) \right) }{\xi} \left( 1-\frac{\xi}{R_i} \right) \ ,
\eeqn
Note that the values of the solute concentrations appearing in these expressions may also change due to the perturbation, according to the constraints in the system. If $\Joi > \Jio$ the drop further expand, accentuating the initial perturbation and the system is therefore unstable against Ostwald ripening. On the contrary if $\Joi < \Jio$ the drop shrinks back, implying that the system is stable. We will solve this problem by performing a linear stability analysis in the small parameter $\epsilon/R^*$.

We first derive the $\epsilon$ dependency of the concentrations appearing in the flux expressions (\eqs \eqref{app:Joi_large} and \eqref{app:Jio_large}). Since the Gibbs-Thomson relations dictate the interface concentrations of the solute (constraint (i) in \ref{sec:app:large:steady}) we find:
\beqn
\Pin(R_i) &=& \Pin(R^*) \\
\Po(R_i) &=& \Po(R^*) - \frac{\hatPo l_c }{(R^*)^2} \epsilon + \mathcal{O}((\epsilon/R^*)^2) \ .
\eeqn
Then, enforcing the constraints of chemical equilibrium far from the interface, and of homogeneous combined concentrations $P+S$ both inside and outside the drop (constraints (ii) and (iv) respectively, \ref{sec:app:large:steady}), we find
\beqn
\Pin(0) &=& \Pin^*(0) +  \frac{\hatPo l_c }{(R^*)^2} \epsilon + \mathcal{O}((\epsilon/R^*)^2) \ ,
\eeqn
where $\Pin^*(0)$ is the solute concentration in the drop centre at the steady state ($R_i=R^*$). To obtain this result we used the fact that the far-field concentrations $\Pinf, \So(\infty)$ are not affected by the perturbation. Indeed, the random perturbation of the drop radii leaves the total drop volume unchanged. Then, the interface solute concentrations are perturbed according to the Gibbs-Thomson relations (\GT), thereby modifying the profiles near drops. However since the solute interface concentrations are randomly perturbed (up and down), we expect that the far-field profiles remain unperturbed up to the first order in the perturbation. We also used $k\ll h$ since $k$ must be smaller than $k_u$ (\eq \eqref{app:ku}) in the large drop regime.

We can now express the fluxes as functions of the perturbation :
\beqn
\Joi = J^* + \frac{D}{\xi (R^*)^2} \left( \hatPo l_c - \frac{k \hatPin \xi}{2 h} \right) \epsilon + \mathcal{O}(\delta) \\
\Jio = J^* - \frac{D}{\xi (R^*)^2} \left( \hatPo l_c - \frac{k \hatPin \xi}{2 h} \right) \epsilon +  \mathcal{O}(\delta)
\eeqn
with 
\beqn
J^* &\equiv& \Joi(R^*)=\Jio(R^*) \\
\delta  &=& \mathcal{O}((\epsilon/R^*)^2) + \mathcal{O}((\epsilon/R^*)(\xi/R^*)(\hatPo l_c/R^*)) \ ,
\eeqn
To obtain this expression we used the results from the steady state (\ref{sec:app:large:steady}) that lead to $\Pinf - \Po(R^*) \approx \Pin(R^*) - \Pin^*(0) \approx k \hatPin/2$. At small forward rate constant $k$ the term $\hatPo l_c$ originating from the Gibbs-Thomson relation (\eq \eqref{gtP}) dominates over the chemical reaction induced-term $k\hatPin\xi/(2h)$. This implies the influx $\Joi$ being larger than the efflux $\Jio$, and therefore drop growth. In this scenario the perturbation is amplified, and the system is therefore unstable against Ostwald ripening. On the contrary when $k$ is large enough we have $\Joi<\Jio$, leading to drop shrinkage and the multi-drop system is stable. This confirms our intuitive arguments in \Sec \ref{sec:large} that chemical reactions tend to stabilise a multi-drop system. The critical rate $k_l$ of the transition between the unstable and stable regime is the solution of $\hatPo l_c =  k_l \hatPin \xi/(2h)$, hence \eq \eqref{k_l}.

\ack
We thank Pablo Sartori, David Schnoerr and Tom Leyshon for useful comments on the manuscript.

\section*{References}

\bibliographystyle{unsrt}

\providecommand{\newblock}{}

\end{document}